\documentclass[fleqn, usenatbib]{mnras}

\usepackage{newtxtext,newtxmath}

\usepackage[T1]{fontenc}

\DeclareRobustCommand{\VAN}[3]{#2}
\let\VANthebibliography\thebibliography
\def\thebibliography{\DeclareRobustCommand{\VAN}[3]{##3}\VANthebibliography}

\usepackage{orcidlink}
\usepackage{graphicx}	

\title[Radio and winds in BALQSOs]{Connecting radio emission to AGN wind properties with Broad Absorption Line Quasars}

\author[J. Petley]{
J. W. Petley,$^{1}$\thanks{E-mail: james.w.petley@durham.ac.uk}\orcidlink{0000-0002-4496-0754}
L. K. Morabito,$^{1,2}$\orcidlink{0000-0003-0487-6651}
D. M. Alexander,$^{1}$
A. L. Rankine,$^{3}$
V. A. Fawcett,$^{1}$
D. J. Rosario,$^{4}$\newauthor
J. H. Matthews,$^{5}$
T. M. Shimwell,$^{6, 7}$
A. Drabent$^{8}$
\\
$^{1}$Centre for Extragalactic Astronomy, Department of Physics, Durham University, Durham, DH1 3LE, UK\\
$^{2}$Institute for Computational Cosmology, Department of Physics, Durham University, Durham, DH1 3LE, UK\\
$^{3}$Institute for Astronomy, University of Edinburgh, Royal Observatory, Edinburgh, EH9 3HJ, UK\\
$^{4}$School of Mathematics, Statistics and Physics, Newcastle University, NE1 7RU, UK\\
$^{5}$Institute of Astronomy, University of Cambridge, Madingley Road, Cambridge, CB3 0HA\\
$^{6}$ASTRON, Netherlands Institute for Radio Astronomy, Oude Hoogeveensedijk 4, 7991 PD, Dwingeloo, The Netherlands\\
$^{7}$Leiden Observatory, Leiden University, PO Box 9513, NL-2300 RA Leiden, The Netherlands\\
$^{8}$Thuringer Landessternwarte, Sternwarte 5, D-07778 Tautenburg, Germany\\
}

\date{Accepted XXX. Received YYY; in original form ZZZ}

\pubyear{2022}

\begin{document}
\label{firstpage}
\pagerange{\pageref{firstpage}--\pageref{lastpage}}
\maketitle

\begin{abstract}
Broad Absorption Line Quasars (BALQSOs) show strong signatures of powerful outflows, with the potential to alter the cosmic history of their host galaxies. These signatures are only seen in  $~\sim$10\% of optically selected quasars, although the fraction significantly increases in IR and radio selected samples. A proven physical explanation for this observed fraction has yet to be found, along with a determination of why this fraction increases at radio wavelengths. We present the largest sample of radio matched BALQSOs using the LOFAR Two-metre Sky Survey Data Release 2 and employ it to investigate radio properties of BALQSOs. Within the DR2 footprint, there are 3537 BALQSOs from Sloan Digital Sky Survey DR12 with continuum signal to noise $\ge 5$. We find radio-detections for 1108 BALQSOs, with an important sub-population of 120 LoBALs, an unprecedented sample size for radio matched BALQSOs given the sky coverage to date. BALQSOs are a radio-quiet population that show an increase of $\times 1.50$ radio-detection fraction compared to non-BALQSOs. LoBALs show an increase of $\times 2.22$ that of non-BALQSO quasars. We show that this detection fraction correlates with wind-strength, reddening and \ion{C}{iv} emission properties of BALQSOs and that these features may be connected, although no single property can fully explain the enhanced radio detection fraction. 
We create composite spectra for sub-classes of BALQSOs based on wind strength and colour, finding differences in the absorption profiles of radio-detected and radio-undetected sources, particularly for LoBALs. Overall, we favour a wind-ISM interaction explanation for the increased radio-detection fraction of BALQSOs.
\end{abstract}




\section{Introduction}

The importance of black holes in overall cosmic history seems to grow with every paper and new discovery in the field. The key confirmed result that demonstrates black holes are not simply extreme isolated phenomena is the black hole mass-stellar velocity dispersion correlation \citep{Ferrarese2000AGalaxies, Gebhardt2000ADispersion, Kormendy2013CoevolutionGalaxies}. This immediately ties together the growth of black holes and their host galaxies, and raises many questions about what possible mechanisms could connect them. We now also understand through cosmological simulations that star formation rates need to be moderated by feedback mechanisms, usually involving black holes, for our simulations to align with the observed star formation history of the universe \citep{Bower2006BreakingFormation, Croton2006TheGalaxies}. Many galaxies have a compact central region emitting much more radiation than can explained by star formation in the same region. This non-stellar radiation is explained by accretion around supermassive black holes in these compact regions and the phenomena as a whole are refered to as Active Galactic Nuclei (AGN). The idea of AGN feedback has arisen as an explanation in which accretion processes at the smallest scales of a galaxy can have a large influence over its evolution for billions of years. 

Quasars (also QSOs) are the most powerful class of observed AGN. The subclass of quasars known as Broad Absorption Line Quasars (BALQSOs) are some of the most important candidates for observing AGN $\rightarrow$ Host Galaxy feedback mechanisms as they show strong blue-shifted absorption troughs in the spectra due to fast moving outflowing gas intersecting our line of sight. The rest of the QSO population does not show these distinctive absorption features. Since BALQSOs were first observed \citep{Foltz1987}, and their outflows defined and quantified \citep{Weymann1991}, much research has gone into probing how these sources differ to the general quasar population across all wavelengths. These investigations, especially at high and low (X-ray and radio) frequencies, are often limited by small number statistics as BALQSOs constitute only $\sim$10\% of optically selected quasars \citep{Trump2006, Gibson2009}.

The presence a particular fraction of quasars displaying these outflow signatures has been explained through two primary theories. One hypothesis is an orientation approach where changing wind signatures are produced by different viewing angles of the quasar through the broad line region \citep{Weymann1991, Ghosh2007a}. The other is an evolutionary model where the observed fraction of BALQSOs is a measure of the lifetime of a BALQSO phase. This phase is usually postulated to be early in the life of the quasar due to connections with reddening and proposed mechanisms for removing material from the host galaxy \citep{Lipari2009}. 

At typical redshifts for probing the height of cosmic star formation ($1.5<z<4$), see review of \cite{Madau2014CosmicHistory}, the lines used to identify BALQSOs fall in the ultra-violet (UV) range covered by large scale spectroscopic surveys such as the Sloan Digital Sky Survey \citep[SDSS -][]{York2000}. All BALQSOs show high ionisation absorption features, particularly \ion{C}{iv} (1550 $\Angstrom$), which provides the initial classification. BALQSOs with similar broad absorption features at \ion{Mg}{ii} or \ion{Al}{iii} are known as Low Ionisation BALQSOs (LoBALs). LoBALs are otherwise largely similar to HiBALs \citep{Schulze20172.5} but have some key differences. Compared to HiBALs, LoBALs show higher gas column densities and velocities \citep{Hamann2019OnOutflows}, they have a greater fraction in dust reddened samples \citep{Urrutia2009TheQuasars} and they are fainter in the X-ray \citep{Green2001AQuasars}. LoBALs account for around 10\% of BALQSOs, the other 90\% being HiBALs, and therefore have only seen limited study in large sample sizes. The relationship between HiBALs and LoBALs is unclear but could be interpreted in the wider context of the BALQSO orientation and evolution hypothesis mentioned previously. In the orientation model, LoBALs could be a subset of the covering angle for BALQSOs where sufficient shielding from the radiation allows for the lower ionisation species to be observed in absorption. Others, however, place LoBALs in an evolutionary framework, usually in an earlier stage in the life cycle of the quasar than HiBALs due to features such as their increased reddening. 

Many studies of the observable properties of BALQSOs, such as their UV continuum, as well as many derived parameters, such as their black hole masses, have found BALQSOs to be indistinguishable from the general quasar population (Black Hole Mass: \cite{Schulze20172.5}; UV: \cite{Dipompeo2012TheQuasars}). This is often interpreted as strong support for the orientation interpretation as many features do not point towards a special evolutionary phase. However, there are some key differences of note which in many cases become further exaggerated in LoBALs. 

A key feature of the UV spectra of BALQSOs is their significant reddening, particularly for LoBALs, which has generally been attributed to dust attenuation \citep{Sprayberry1992ExtinctionObjects, Reichard2003ARelease}. This means that on average a BALQSO will need to be more luminous to be detected down to a given optical flux limit than an average QSO which also may cause underestimation of the intrinsic fraction of BALQSOs \citep{Dai20072MASSBALQSOs, Allen2010AFraction} when using optically selected surveys and has impact when using quasar SED templates in various aspects of analysis, a common practice in luminosity corrections and extrapolation of physical parameters such as star formation rate. BALQSOs are also intrinsically X-ray weak compared to non-BALQSOs \citep{Clavel2005AXMM-Newton, Morabito2014}, a property which is not orientation dependent and means X-rays are less important to the ionisation state of the outflowing BAL wind.

Radio wavelength studies of BALQSOs are of key importance since radio emission can be traced all the way from the dynamics of the black hole to the whole scale of the galaxy depending on the emission mechanism. Also radio emission is not absorbed by intervening dust and can therefore provide an orientation-independent measurement of a radio emitting process, apart from in rare cases of relativistic beaming. Classically, radio populations have been split into \textit{radio-loud} and \textit{radio-quiet} due to the apparent bi-modality of radio galaxies when looking at the ratio between their radio and optical luminosity \citep{Kellermann1989}.  Early and subsequent studies have firmly placed the BALQSO population in the \textit{radio-quiet} regime \citep{Stocke1992TheQSOs, Becker2000PropertiesSurvey, Morabito2019}. While radio-loud emission is generally attributed to large-scale powerful jets which, if resolved, can allow for measurement of size and orientation \citep{Barthel1989IsBeamed}, radio-quiet sources have several possible emission mechanisms such as frustrated jets, wind shock emission and star formation. While radio-loud BALQSOs do exist and have been studied \citep{Stocke1992TheQSOs, Brotherton1998DiscoverySelection}, often historically because they are brighter and easier to observe, they are a much lower fraction of overall radio-detected BALQSOs when compared to the general quasar population. This implies that BALQSOs are less likely to contain large-scale radio jets than the general quasar population.

Although BALQSOs are radio-quiet, much of the previous radio work on these sources has focused on the small fraction which are radio-loud. Surveys such as Faint Images of the Radio Sky at Twenty centimetres (FIRST) \citep{Becker1995TheCentimeters} are more biased towards radio-loud sources and Very Long Baseline Interferometry (VLBI) observations of BALQSOs have focused on radio-loud sources with the aim to resolve and measure the orientation of the expected jet structure in these sources. A key result using FIRST, along with 2MASS \citep{Skrutskie2006The2MASS} and SDSS \citep{Adelman-McCarthy2007TheSurvey}, is that of \cite{Urrutia2009TheQuasars} who found that when they created a sample of reddened, largely radio-loud, quasars there was a high fraction of BALQSOs in the sample, and all but one were LoBALs. This was seen as strong evidence for the evolutionary hypothesis as the reddening is considered a feature of an earlier stage in the life of a quasar as material is removed from the galaxy by feedback. 

Previous VLBI observations of BALQSOs have revealed mixed results that have thus far failed to conclusively support one interpretation over the other. In \cite{Jiang2003EVNQuasars} 3 BALQSOs were observed with the European VLBI Network (EVN) and all sources had jet sizes less than 1kpc. They also observed a flat spectrum source with orientation likely close to the line of sight, which would support the presence of BAL winds close to the jet axis. \cite{Doi2009VLBIQuasars} find a similar result when using the Optically Connected Array for VLBI Exploration (OCTAVE) instrument to observe 22 radio-loud BALQSOs. They find 4 compact inverted spectrum sources allowing for two possible explanations. Either the radio sources are very young, or they are observing Doppler boosted emission along the line of the jet. In either case, those observations cannot be explained by the classic equatorial disk wind viewing angle of a standard quasar. In  \cite{Bruni2013TheQuasars} 9 BALQSOs were studied with the EVN and Very Long Baseline Array (VLBA). They found a variety of structures with jet sizes from tens of parsecs to a hundreds of kiloparsecs. They found this supported the geometric interpretation where the sizes relate to different potential viewing angles of the wind with respect to the jet axis and that they could occur throughout quasar evolution allowing for jet sizes to grow to a range of sizes. \cite{Cegowski2015VLBI0} isolated BALQSOs with weaker wind strengths and observed them with the EVN and VLBA. They find that these sources are on average potentially younger than other BALQSOs and are more radio-loud. However, the radio-loudness could be associated with different levels of absorption, also allowing for a geometric interpretation. All of these studies focused on radio-loud sources which are the exception to most BALQSOs. Although jets were observed in some of the sources it is not known whether this emission mechanism continues down to the radio-quiet sources or whether other mechanisms become the dominant component.

Other radio studies of BALQSOs aiming to use large sampels have been limited to higher frequency surveys such as NVSS \citep{Condon1998TheSurvey} and FIRST because these have been the most sensitive large area surveys available. This limits the sensitivity to the synchrotron emission which increases in luminosity at longer wavelengths. Therefore higher frequency surveys pick out more radio-loud sources which are not representative of the overall population. With the development of LOFAR \citep{vanHaarlem2013LOFAR:ARray} we now have access to increased sensitivity at a resolution matched to that of FIRST, but at an order of magnitude lower sensitivity. As well as increased source counts, the lower frequency emission from quasars is dominated by optically thin synchotron emission. Due to the shape of the synchotron spectrum, the radio emission is brighter at lower frequencies when compared to the flat spectrum of radio emission from optically thick regimes which have a higher relative contribution at $\sim$ GHz frequencies. 

In this paper we inspect and expand upon the work of \cite{Morabito2019} which utilised the LOFAR Two Metre Sky Survey (LoTSS) Data Release 1 \citep{Shimwell2019TheRelease}. In that work BALQSOs were found to have a detection fraction independent of luminosity, a result that is not seen at higher frequencies. Interestingly it was also tentatively found that detection fraction did correlate with absorption properties. This shows that the radio emission in BALQSOs is tied to the current state of the BAL wind and that structural models for BALQSOs need to be able to also explain the radio properties of these sources. 

We confirm some of the results from \cite{Morabito2019} using LoTSS DR2 \citep{Shimwell2022TheRelease} which spans 5740 square degrees, or 13 times the area of LoTSS-DR1, with a median sensitivity of 83~$\mu$Jy beam$^{-1}$ at 144~MHz. This has allowed us to expand our statistical analysis of BALQSOs, particularly the rarer LoBALs which we can split out more easily, but also maintain high spectra clarity and avoiding some biases in \cite{Morabito2019} by introducing UV continuum signal to noise cuts while still creating a larger sample. We explore the potential explanations of the link between radio emission and wind strength by developing simple models for different geometries and evolution and testing them against our large sample. We also identify the potential focuses for future studies of BALQSOs .  

Throughout this paper we assume a cosmology defined by the parameters determined by the \textbf{Planck 2015} observations and published in \cite{Ade2016}. This is a standard, flat $\Lambda$CDM cosmology with $H_0$= 67.8 km s$^{-1}$ Mpc$^{-1}$ and $\Omega_M$ = 0.308. We use the spectral index convention of $S_\nu \propto \nu^{-\alpha}$.

\section{Data and Methods}

\subsection{Radio Data}

The core data dataset for this work is the LOFAR Two Metre Sky Survey (LoTSS) Data Release 2 \citep{Shimwell2022TheRelease}\footnote{\url{https://lofar-surveys.org/surveys.html}}. This release of the 144~MHz survey covers 5740 square degrees across two separated fields with a with a median sensitivity of 83~$\mu$Jy beam$^{-1}$ at the central frequency. There are 4,396,228 sources above a detection limit of $5 \sigma$. At these frequencies, for most sources, we expect synchotron emission to be the dominant contribution to the observed flux; the flux from this spectral slope increases as frequency decreases which reaches a fainter population when compared to surveys of similar sensitivity but at higher frequency. 

There is a preliminary optical cross-matched catalogue in use based on automatic matching of simple structure radio sources to optical images. However, this system struggles on more complex radio structure. At the time of writing, there is an ongoing citizen science project entitled the \textit{"LOFAR Galaxy Zoo"}\footnote{\url{https://www.zooniverse.org/projects/chrismrp/radio-galaxy-zoo-lofar}} taking place on the \textit{Zooniverse} website. This project allows the public to help characterise more complicated extended emission systems that cannot be identified automatically, and also allows users to identify potential optical counterparts. Sources currently being processed by the citizen science project have not yet been added to the cross matched catalogue that was used for this paper. It is unlikely that the inclusion of sources in the LOFAR Galaxy Zoo would change the results of this paper significantly as they largely concern complex structure radio-loud sources which do not account for a large population of BALQSOs.

\subsection{Balnicity Index and radio-loudness}
  
Without knowledge of the exact geometry and energetics of BALQSO systems, the identification and characterisation of outflows in BALQSOs is typically uses the Balnicity Index (BI) defined in \cite{Weymann1991} at the \ion{C}{iv}. This can be thought of as some sort of proxy for the "wind strength" and has been used to compare BALQSO absorption properties. This index uses the depth and breadth of the absorption trough and is defined 

\begin{equation}
\verb|BI| = - \int^{3000}_{25000} [1 - \frac{F(v)}{0.9}] C_B dv  
\end{equation}

\noindent $v$ is the velocity in km~s$^{-1}$ with respect to the line centre. $F(v)$ is the flux after the removal of the QSO continuum, obtained by a fit to the original spectra, and $C_B$ is a constant which is set to 1 if $F(v)$ is lower than 0.9 for 2000 km~s$^{-1}$ and 0 otherwise. This minimum velocity and absorption required for the integration to begin is set so that intervening systems in the absorption trough are removed and that the absorption is definitely broad. The continuum normalised flux is created by a fitting process which extracts the QSO continuum emission from spectra. Then, by dividing the whole spectra by that continuum, the genuine continuum emission will be normalised to one. Therefore a value of $F(v)$ of 0.9 means that the flux at that frequency is at 90\% that of the continuum fit at the same frequency.  Although the standard practice is to define BALQSOs as all sources with BI~>~0, there are some well known considerations to be made when using this definition. Alternative outflow strength indices exist, such as the Absorption Index (AI; \citealt{Hall2002UnusualSurvey, Trump2006}), which do not require a minimum velocity in the integration limits. This means that the equation is identical to that above but the upper limit of the integral is 0 km~s${-1}$ rather then 3000 km~s${-1}$. Therefore using AI to classify BALQSOs will create larger, but potentially more contaminated, samples as visual inspection of sources with AI>0 and BI = 0 includes sources that do not have clear absorption features. A study of sources with positive AI shows a bi-modality in the AI distribution where only the higher AI distribution contains genuine BALQSOs \citep{Knigge2008TheQuasars}. More advanced selection criteria can be used such as various machine learning techniques. However, these methods introduce their own uncertainties and contaminations. Rather than potentially contaminate the sample we use the classic BI definition which ensures all of our BALQSOs show clear features at the \ion{C}{iv} line line at the cost of missing some genuine BALQSOs.

\begin{figure}
    \centering
    \includegraphics[width=0.9\linewidth]{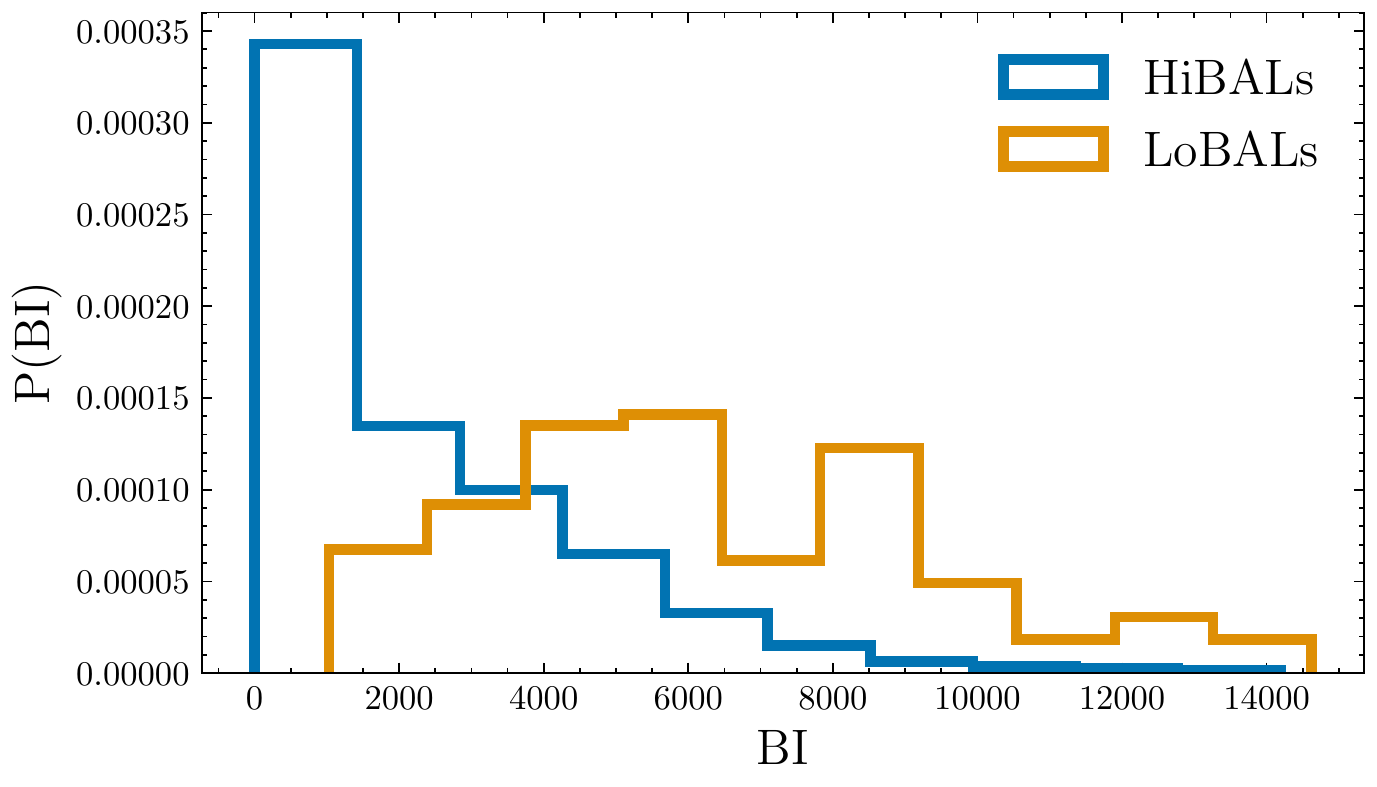}
    \caption{Balnicity index (BI) distributions for HiBALs (blue) and LoBALs (orange). Typically LoBALs have larger BI values than HiBALs, as measured from the \ion{C}{iv} absorption.}
    \label{fig:BI_dist}
\end{figure}

\autoref{fig:BI_dist} displays the BI distribution of our final LoBAL and HiBAL sample; it is clear the LoBALs tend to have/occupy higher BI values compared to the HiBALs, which is consistent with previous observational studies \citep{Reichard2003ARelease, Morabito2019} and simulations \citep{Higginbottom2013AQuasars, Matthews2016TestingWinds, Matthews2020StratifiedProperties}. In fact, the high BI end (>~10,000 km~s$^{-1}$) of our sample is entirely dominated by LoBALs.

Radio-loudness is a concept that has been used to characterise whether a source shows an excess in radio emission when compared to optical \citep{Kellermann1989}. We define radio-loudness as the ratio ($R$) between the 144~MHz luminosity and the B-band ($\sim 450~nm$) rest-frame luminosity. We use a similar approach to \cite{Morabito2019} which we briefly describe below. To obtain this optical luminosity we take the 3000~$\Angstrom$ luminosity from SDSS DR12 and convert it to a flux based on the spectroscopic SDSS redshift of the source. We then correct this flux for extinction within the Milky Way (using \cite{Pei1992InterstellarClouds}) and correct the rest frame emission for intrinsic extinction from the host galaxy using a Small Magellanic Cloud (SMC) model which was found to be a good fit for BALQSOs by \cite{Reichard2003ARelease}. We finally shift this rest frame emission to the B-band using the mean spectral energy distribution (SED) for Type 1 Quasars from \cite{Richards2006}. A correction is also applied when calculating the radio luminosity from the LoTSS total flux, assuming a synchotron spectral index of $\alpha = 0.7$. This gives us a consistent measure for radio-loudness across the whole sample under the assumption that the SED and extinction curves for BALQSOs and non-BALQSOs are the same (see discussion of BALQSO wind-radio connection).

\begin{figure}
    \centering
    \includegraphics[width = 0.9\linewidth]{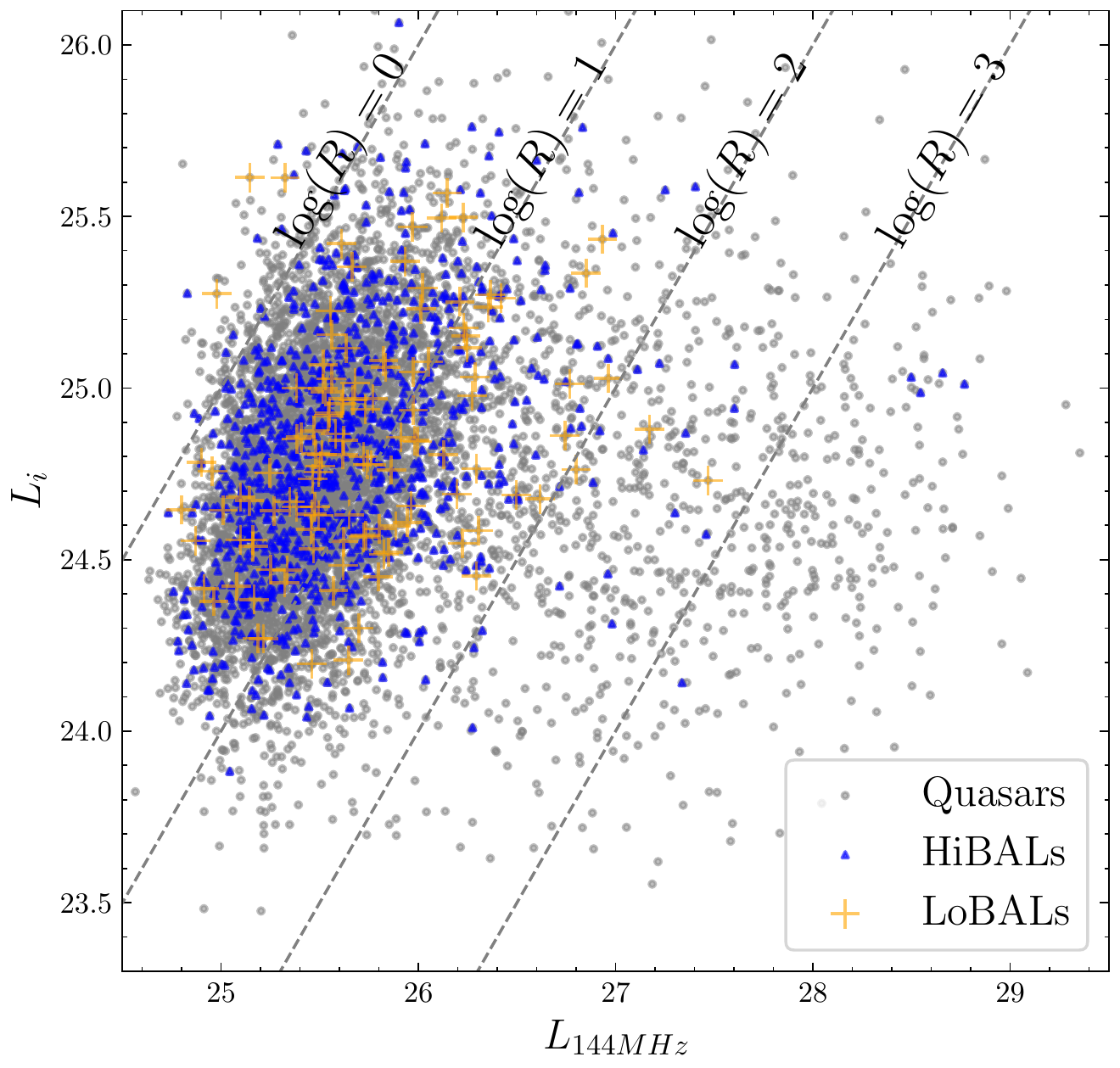}
    \caption{$i$ band (806~nm) luminosity against LoTSS 144MHz luminosity for the full sample of radio-matched quasars with HiBALs in blue and LoBALs in orange. We overlay lines of constant radio-loudness to show to what loudness level our sample is complete.}
    \label{fig:complete_loudness}
\end{figure}

We investigate the radio-loudness to which our sample is complete as this is important in understanding what populations we are able to properly characterise depending on their radio-loudness. In \autoref{fig:complete_loudness} we show that our combined spectroscopic and radio sample is largely complete to a radio-loudness close to $\log(R) = 0$.

Typically, the general quasar population can be split into a radio-loud and radio-quiet population. This can be seen in the distribution of these sources in \autoref{fig:complete_loudness}.  The total distribution appears to be the result of two overlapping distributions with a border around $\log(R) = 1.5$. The majority of the population lies above and to the left of $\log(R) = 1.5$ with outliers below and to the right in \autoref{fig:complete_loudness}. The population drops in number quickly in the radio-loud regime, especially for LoBALs. Previous radio studies investigating these distributions have often operated around 5~GHz frequencies and the radio-loud distinction occurs around $R = 10$. The border value we find at 144~MHz corresponds very closely to a scaling of the traditional value using the expected synchotron spectral index of $\alpha = 0.7$, defined so that in positive direction of spectral index as frequency increases the flux decreases. It is consistent with \cite{Morabito2019}, who used $g$ band rather than $i$ band and thus found a slightly higher value of log(R) as the division between radio-loud and radio-quiet.

We provide the reader with the overall catalogue of LoTSS detected SDSS DR12 quasars in the supplementary material for this work. A summary is provided in \autoref{tab:catalogue_summary}. The redshift range cut to the wavelengths that BALQSOs can be identified and also continuum SNR cuts have not been applied to the data.

\begin{table*}
    \centering
    \begin{tabular}{c|c|c|l}
    \hline
Column & Name & Format & Description  \\
    \hline
    1   & Source\_Name & STRING & LoTSS DR2 Name \\
    2 & RA & DOUBLE & LoTSS DR2 J2000 Right Ascension (degrees) \\
    3 & E\_RA & DOUBLE & LoTSS DR2 J2000 Right Ascension uncertainty (arcseconds) \\
    4 & DEC & DOUBLE & LoTSS DR2 J2000 Declination (degrees) \\
    5 & E\_DEC & DOUBLE & LoTSS DR2 J2000 Declination uncertainty (arcseconds) \\
    6 & Peak\_flux & DOUBLE & LoTSS DR2 144MHz peak flux (mJy/beam) \\
    7 & E\_Peak\_flux & DOUBLE & LoTSS DR2 144MHz peak flux uncertainty (mJy/beam) \\ 
    8 & Total\_flux & DOUBLE & LoTSS DR2 144MHz integrated flux (mJy) \\
    9 & E\_Total\_flux & DOUBLE & LoTSS DR2 144MHz integrated flux uncertainty (mJy) \\ 
    10 & Maj & DOUBLE & LoTSS DR2 major axis (arcseconds) \\
    11 & E\_Maj & DOUBLE & LoTSS DR2 major axis uncertainty (arcseconds) \\
    12 & DC\_Maj & DOUBLE & LoTSS DR2 deconvolved major axis (arcseconds) \\
    13 & E\_DC\_Maj & DOUBLE & LoTSS DR2 deconvolved major axis uncertainty (arcseconds) \\
    \hline 
    14 & SDSS\_NAME & STRING & SDSS DR12Q Name \\
    15 & RA\_1 & DOUBLE & SDSS J200 Right Ascension (degrees) \\ 
    16 & DEC\_1 & DOUBLE & SDSS J200 Declination (degrees) \\
    17 & THING\_ID & INT32 & Unique SDSS identifier \\
    18 & PLATE & INT32 & Spectroscopic plate number \\
    19 & MJD & INT32 & Spectroscopic MJD \\ 
    20 & FIBERID & INT32 & Spectroscopic fiber ID \\
    21 & Z\_VI & DOUBLE & Visually inspected redshift \\
    22 & SNR\_1700 & DOUBLE & Median signal-to-noise ratio per pixel in the window $1650 - 1750\Angstrom$ \\
    23 & BAL\_FLAG\_VI & SHORT & BAL Flag from visual inspection \\ 
    24 & BI\_CIV & DOUBLE & Balnicity index of \ion{C}{iv} trough (km s$^{-1}$) \\
    25 & ERR\_BI\_CIV & DOUBLE & Balnicity index uncertainty of \ion{C}{iv} trough (km s$^{-1}$) \\
    26 & REW\_ALIII & DOUBLE & Rest frame equivalent width of the \ion{Al}{iii} trough ($\Angstrom$) \\
    \hline 
    27 & L3000 & FLOAT & Estimated $3000\Angstrom$ luminosity ($10^{-7} W$) \\
    28 & MBHMgII & FLOAT & Black hole mass estimated from \ion{Mg}{ii} (M$_\odot$) \\ 
    29 & MBHCIV & FLOAT & Black hole mass estimated from \ion{C}{iv} (M$_\odot$) \\
    30 & Lbol & FLOAT & Quasar bolometric luminosity ($10^{-7} W$) \\
    31 & e\_Lbol & FLOAT & Quasar bolometric luminosity uncertainty ($10^{-7} W$) \\
    32 & nEdd & FLOAT & log(Eddington Fraction) \\
    \hline 
    33 & RAJ2000 & DOUBLE & FIRST Right Ascension (degrees) \\ 
    34 & DEJ2000 & DOUBLE & FIRST Declination (degrees) \\
    35 & Fpeak & DOUBLE & FIRST peak flux (mJy/beam) \\
    36 & Fint & DOUBLE & FIRST integrated flux (mJy) \\ 
    37 & Rms & FLOAT & FIRST rms noise around source (mJy) \\
    38 & Seperation & DOUBLE & LoTSS-FIRST Source seperation (arcseconds) \\ 
    \hline 
    39 & REST\_B & DOUBLE & Estimated rest $b$ band luminosity (W) - Values are 0 when \cite{Kozlowski2017} data is missing \\ 
    40 & Radio\_luminosity & DOUBLE &  log(estimated 144MHz rest frame luminosity assuming $\alpha = -0.7$) (W) \\
    41 & Radio\_loudness & DOUBLE & log(Radio/B\_BAND radio loudness) - inf value when REST\_B = 0 
    
    \end{tabular}
    \caption{Supplementary table column description - Horizontal lines denote sections from different source catalogues. From top to bottom they are: LoTSS DR2 \citep{Shimwell2022TheRelease}, SDSS DR12 Quasars \citep{Paris2017VizieR2017}, \protect\cite{Kozlowski2017} quasar parameters, FIRST \citep{Becker1995TheCentimeters} and finally some derived values described in this section. The full table can be accessed online through the supplementary material. The table does not have a signal-to-noise or redshift cut applied.}
    \label{tab:catalogue_summary}
\end{table*}

\subsection{Spectra}

The spectroscopic data needed to identify BALQSOs is taken from the Sloan Digital Sky Survey (SDSS) DR12 quasar catalogue \citep{Eisenstein2011SDSS-III:Systems,Alam2015}. This contains $297301$ QSO spectra along with auxiliary information including BALQSO classification, \ion{C}{iv} Balnicity Index (BI) and Absorption Index (AI) (see Section 3.1) along with emission width information for selected other emission lines. 151595 of these quasars lie within the LoTSS DR2 coverage. BALQSOs in SDSS can only be defined with high quality spectra, otherwise noise will overwhelm the absorption region, and within a redshift region where the CIV line can be observed completely; this restricts the catalogue to a redshift range of $1.7<z<4.3$ with a signal to noise ratio cut on the continuum at $1700 \Angstrom$ of \verb|SNR_1700| $>5$. 

This cut is important as SNR can have a large impact on both the BALQSO optical identification fraction in the sample and the BALQSO radio-detection fraction compared to non-BALQSOs. At low SNR BALQSOs are harder to identify in the SDSS data and so have a much lower fraction. Also the ones that are identified tend to have a higher bolometric luminosity. These effects to exaggerate the radio-detection fraction of BALQSOs if no SNR cut is taken. On the other hand, higher SNR cuts ($~SNR>20$) select sources with higher bolometric luminosity; for the non-BAL quasars this increases the chance of a source being radio-loud and also radio-detected, while BALQSOs are almost exclusively radio-quiet which then changes the value of non-BALQSOs as a comparison population. As a compromise, we choose SNR $>5$ as a compromise between both of these selection biases and maximising the number of BALQSOs in our sample.

The overall BALQSO/non-BALQSO radio-detection fraction ratio across redshift is 1.93 but with a SNR cut of 5 this drops to 1.45. It remains relatively consistent out to a cut at 20 where the value is 1.24 but at this point we would only have 215 HiBALs and 29 LoBALs within LoTSS coverage, significantly limiting the study of these populations in greater detail. A repeated analysis of the radio detection fraction trends with colour and BI with a SNR cut of 10 showed no change in the extraction of correlations but would severely limit any composite spectra work.  

To define BALQSOs using this catalogue we require visual confirmation via the \verb|BAL_FLAG_VI| and a positive balnicity index (\verb|BI_CIV| $> 0$). We then define LoBALs by using the rest emission width of the Al~III species and require \verb|REW_ALIII| $>0$. This is based on the classical definition of rest emission width, which is used in SDSS DR12, where values greater than zero indicate absorption. A visual inspection of the spectra for a substantial amount (~25\%) of HiBALs and LoBALs, and the full inspection of smaller HiBAL and LoBAL subsets later in this work, confirms this selection criteria is largely successful. We also use the extended catalogue created by \cite{Kozlowski2017} which adds black-hole mass and bolometric luminosity estimates for all of the SDSS DR12 quasars. Finally, we also use the Faint Images of the Radio Sky at Twenty centimetres (FIRST) survey \citep{Becker1995TheCentimeters} which is the best comparative survey at a similar resolution (5 arcsecond) and higher frequency (1.4 GHz). 

After restricting the SDSS data to lie within the Multi-Order Coverage map (MOC) of LoTSS DR2 we match SDSS and the optical counterpart coordinates within the LoTSS DR2 catalogue with a matching radius of 1 arcsecond. Since we are using the optical identifications available in the DR2 catalogue which are based on Pan-STARRS and WISE, this is likely to yield extremely little contamination when matched to SDSS. For a full discussion of the LoTSS optical identification process please refer to \cite{Williams2019TheCatalogue}. This yields 6813 radio-matched QSOs including 988 HiBALs and 120 LoBALs. These are larger numbers than were studied in \cite{Morabito2019} even with the SNR cut that we use in this work. This sample is large enough that the statistical properties of BALQSOs can be studied at a level previously unprecedented due to the scarcity of radio-detected BALQSOs, particularly the rarer LoBAL sub-class.

\begin{table}
    \centering
    \begin{tabular}{|c|c|c|c|}
     & General Quasars  & HiBALs & LoBALs \\ \hline
    Within DR2 MOC   &  31752 & 3270 & 267 \\ 
    radio-detected & 6813 & 1108 & 120 \\
    \end{tabular}
    \caption{Summary of overall sample of SDSS quasars within the LoTSS DR2 coverage.}
    \label{tab:overall_number}
\end{table}

\section{Results}
  
\subsection{Radio and Bolometric Luminosities}

\begin{figure}
    \centering
    \includegraphics[width=8.5cm]{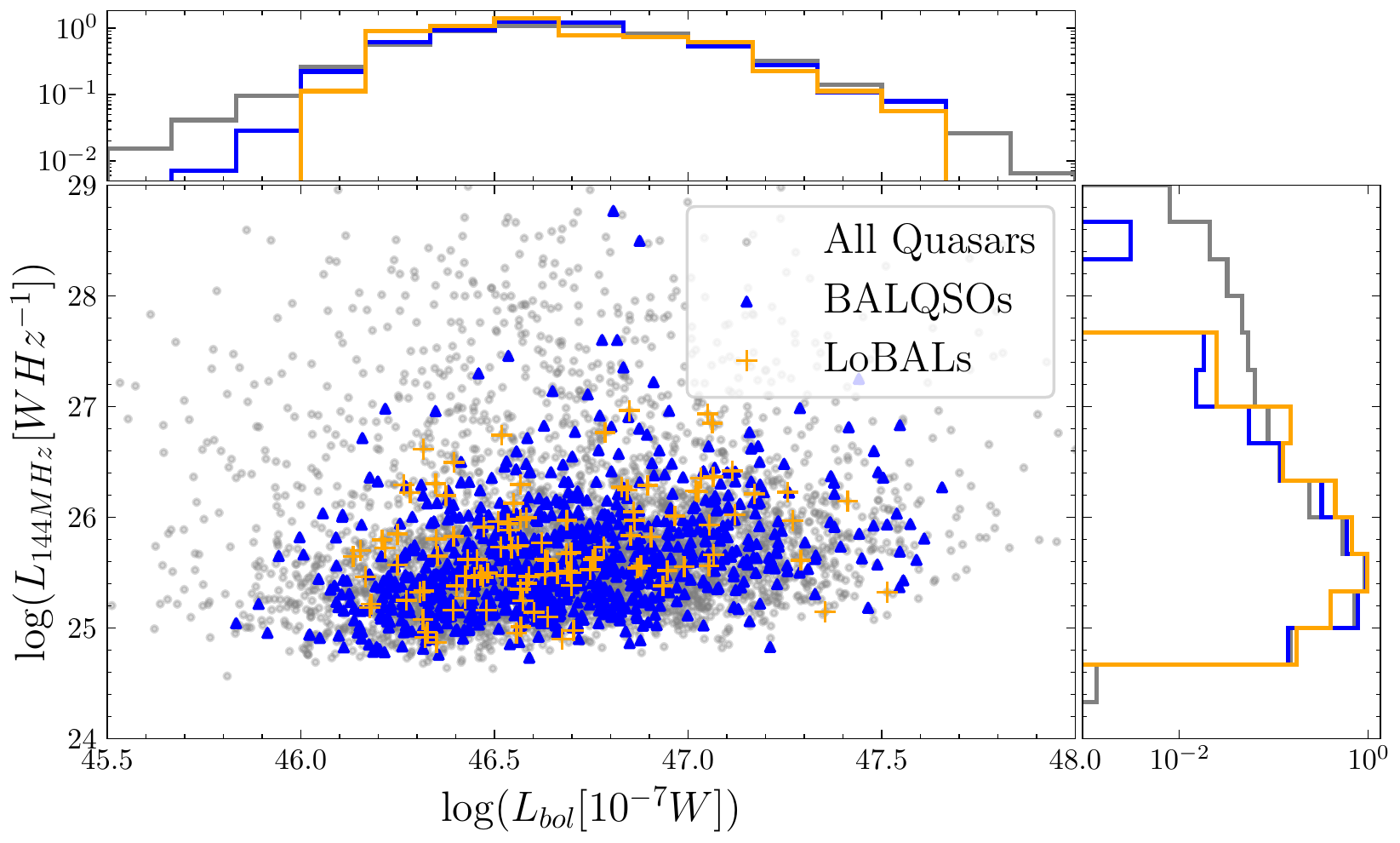}
    \caption{The radio and bolometric luminosity distributions for our HiBAL (blue) and LoBAL (orange) samples. The BALQSO and LoBAL sample do not reach as low bolometric luminosities as the general population due likely due to the effects of reddening. However, this is a small difference and is mirrored in the absence of high radio luminosity BALQSOs and LoBALs. We therefore conclude that differences in radio properties cannot be primarily driven by increased luminosity.}
    \label{fig:radio_vs_bol}
\end{figure}

 Using the bolometric luminosities provided by \cite{Kozlowski2017} we show the distributions of radio luminosity at 144 MHz and bolometric luminosities for the sample in \autoref{fig:radio_vs_bol}. The BALQSOs within the radio matched sample of quasars show very little difference in the distribution of their radio luminosities to the parent population. This is also true for LoBALs. Potentially, LoBALs show a slight increase in bolometric luminosities compared to HiBALs and the rest of the sample but this could be due to the inability to calculate the bolometric luminosity for low luminosity BALQSOs, which are likely to be highly reddened and have lower signal-to-noise spectra.

\begin{figure}
    \centering
    \includegraphics[width=8.5cm]{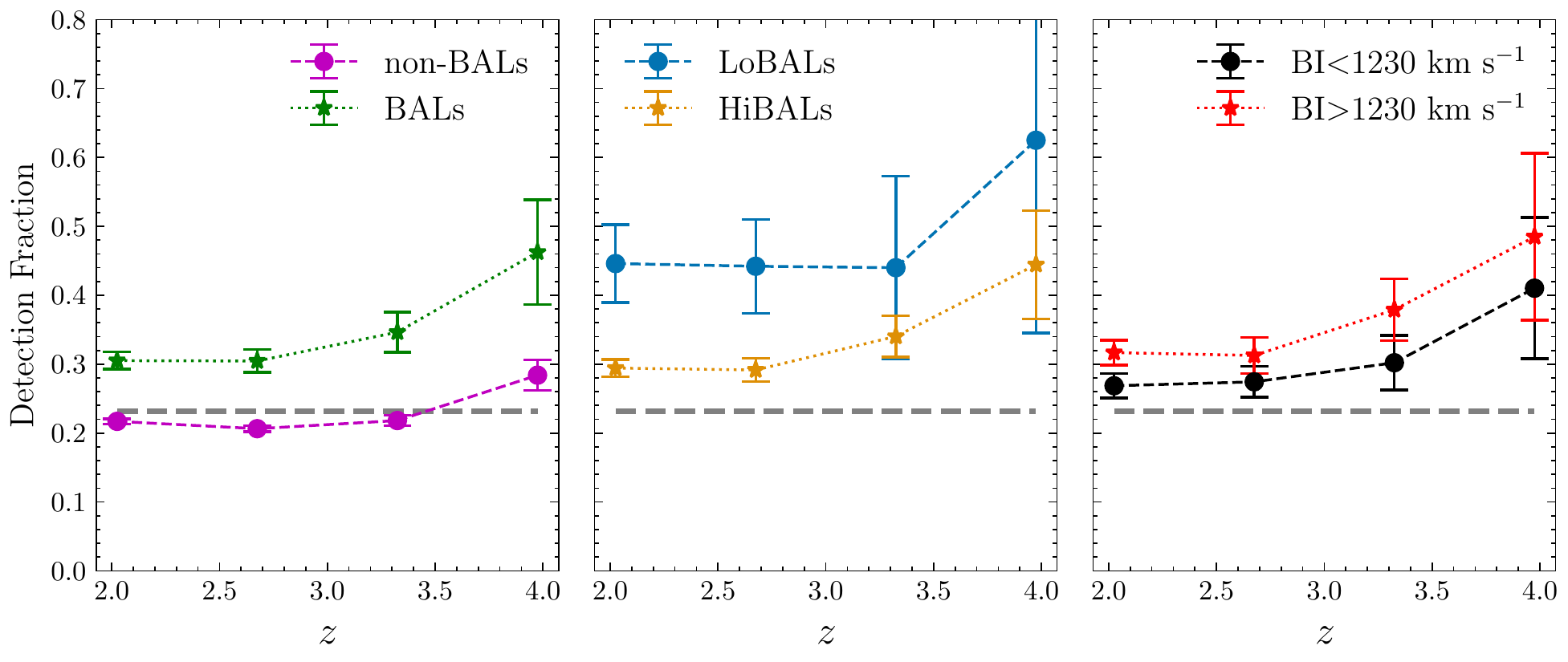}
    \caption{From left to right: the radio-detection fraction for BALQSOs and non-BALQSOs, LoBALs and HiBALs and finally weak and strong HiBALs based on a median BI cut. The uncertainties are estimated through bootstrap sampling of the SDSS sample and repeating the detection fraction calculations. The grey dashed line present in each panel is the mean detection fraction for non-BALQSOs across redshift. We find a dramatic difference in radio-detection fraction between BALQSOs and non-BALQSOs and furthermore between LoBALs and HiBALs. The orange line represents the mean detection fraction for the overall population. This is persistent across redshift and stays relatively constant across redshift. We also present the difference between high BI and low BI sources however it should be noted that the HiBALs and LoBALs also have very different BI distributions (\autoref{fig:BI_dist}).}
    \label{fig:radio_detection_fraction}
\end{figure}

BALQSOs show a difference to the general quasar population in radio-detection fraction, as shown in \autoref{fig:radio_detection_fraction}. We present this in a similar fashion to \cite{Morabito2019}, however we now have the number of sources needed to constrain the uncertainty on detection fraction for HiBALs and LoBALs. We use bootstrap sampling of the whole SDSS DR12 quasar sample and recalculate LoTSS detection fractions to estimate the uncertainty at each point. BALQSOs show $\times 1.5$ the radio-detection fraction when compared to non-BALQSOs across the full redshift range of the sample. The difference only increases further when comparing LoBALs to non-BALQSOs where LoBALs show an even further enhancement in radio-detection fraction and detection rates approaching 50\%. This is the strongest evidence yet that BALQSO winds (in particular LoBAL winds) and radio emission are connected and that this connection spans the era of maximum star formation in cosmic history ($2.5<z<3.5$). This high radio-detection for BALQSOs highlights the importance of creating radio selected samples of quasars when attempting to understand the intrinsic properties of BALQSOs. In the near future this will be best achieved by WEAVE-LOFAR \citep{Smith2016TheSurvey} which should yield a significant number of BALQSO spectra.

In combination with the radio loudness and radio luminosity data presented in \autoref{fig:complete_loudness} and \autoref{fig:radio_vs_bol}, we see that although BALQSOs are more likely to be radio detected than non-BALQSOs that does not necessarily mean they are more luminous. For BALQSOs that are radio detected, they are under represented in the high radio luminosity and radio loud end of these distributions.

\subsection{Detection Fraction Correlations}

\begin{figure}
    \centering
    \includegraphics[width = 0.9\linewidth]{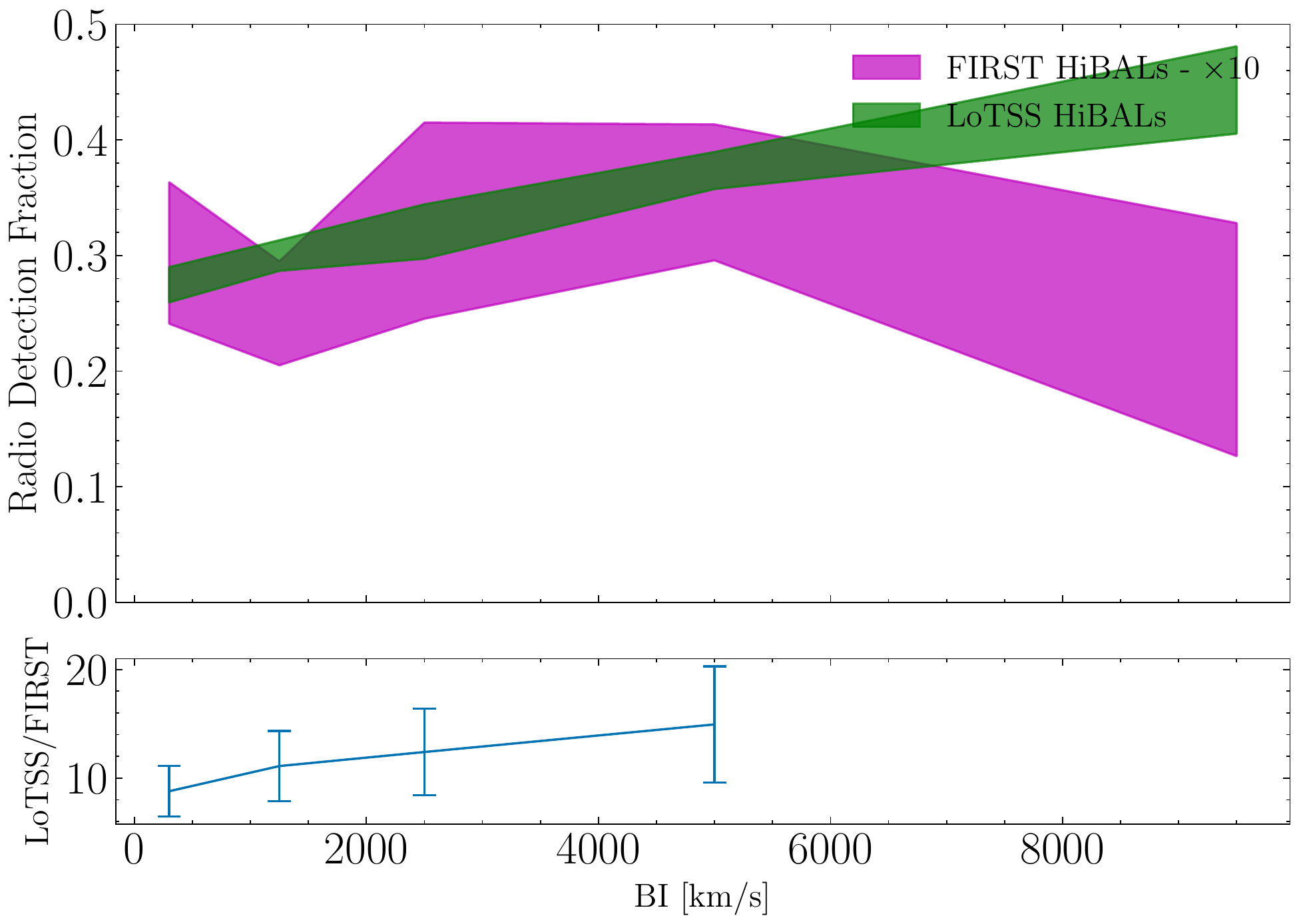}
    \caption{The correlation between BI, a measure of outflow strength, with radio-detection fraction for LoTSS (green) and FIRST (magenta) HiBALs where the first fraction is multiplied by 10 so that the two instruments can be more easily compared. The regions represent the one sigma uncertainty region for the detection fraction calculated by taking bootstrapped samples of the full SDSS DR12 quasar catalogue and calculating the radio-detection fractions each time. The FIRST radio-detection fractions are multiplied by 10 so that they appear on a similar scale. The correlation is only present in LoTSS, the survey that is more sensitive to the radio-quiet population. We find more BALQSOs using LOFAR as BI increases in comparison to FIRST.}
    \label{fig:frac_with_BI}
\end{figure}

We investigate the LoTSS detection fraction further and test its correlation with the main measure of the strength of the outflow, BI. In \autoref{fig:frac_with_BI} we show that the radio-detection fraction does indeed correlate with BI index and also that this relationship is not observed clearly at other radio frequencies such as FIRST. For the BALQSOs in the sample the Pearson correlation coefficient between BI and LoTSS radio-detection fraction is $r = 0.9609$ with a p-value, the probability that a random sample from the population would have an equal to or greater correlation coefficient, of $p=5\times10^{-4}$. For the FIRST results the Pearson coefficient is $r = -0.7$ and the p-value is $p = 0.10$ which is not as strong or significant. This provides a clear link between the appearance of BALQSOs and the low frequency radio emission from their sources.

\begin{figure*}
    \centering
    \includegraphics[width = \textwidth]{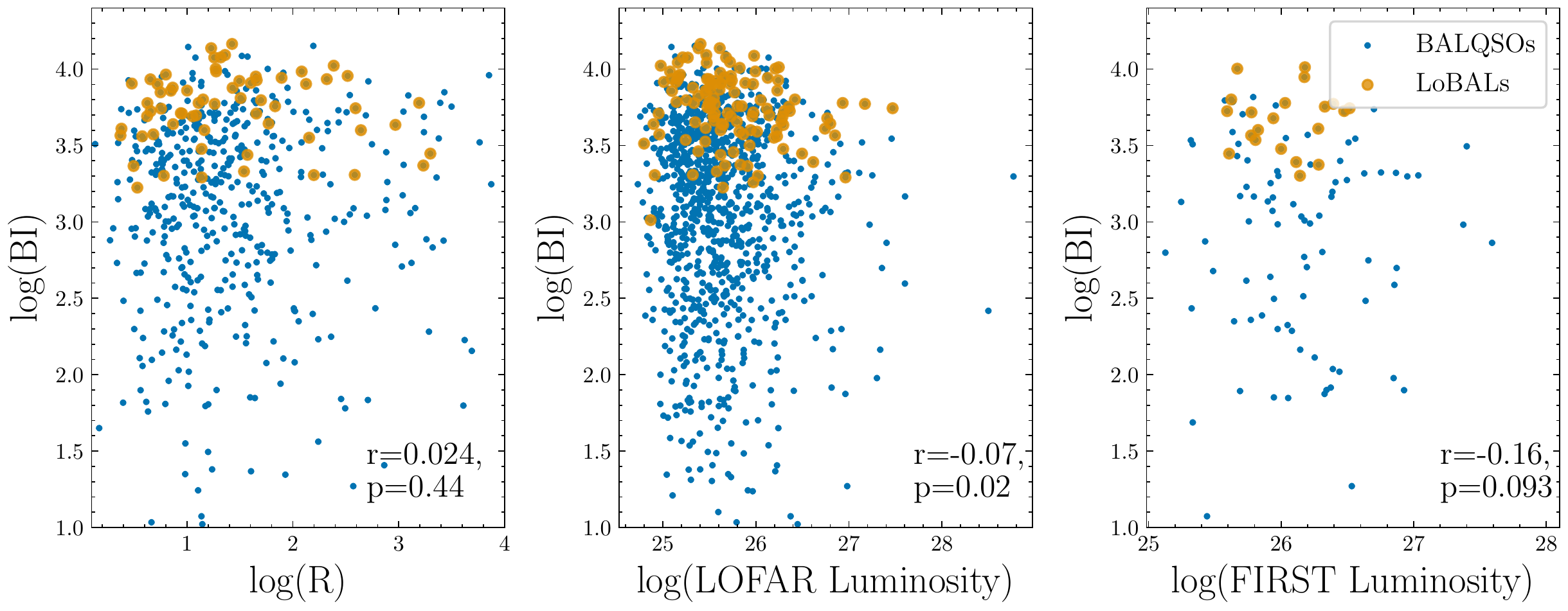}
    \caption{The relation between BI and radio-loudness, LOFAR luminosity and FIRST luminosity for all of the HiBALs (blue) and LoBALs (orange) in our radio matched sample. We display the Pearson correlation coefficient, r, and p-value between the parameters for each figure. We do not find any significant correlations between each of these variables.}
    \label{fig:nonCorrelations}
\end{figure*}

We test whether the increase in LoTSS-detection fraction with BI could be driven by other factors than BI. In \autoref{fig:nonCorrelations} we test whether radio-loudness or luminosity correlate with BI for our sample of BALQSOs. We do not find notable correlations for any of these radio measures leaving only the radio-detection fraction as the key measure that correlates with BI. The Pearson correlation coefficients for BI with radio-loudness, LOFAR luminosity and FIRST luminosity are 0.02, -0.07 and -0.16 respectively with p-values of 0.44, 0.02 and 0.09. 

\begin{figure}
    \centering
    \includegraphics[width = 1\linewidth]{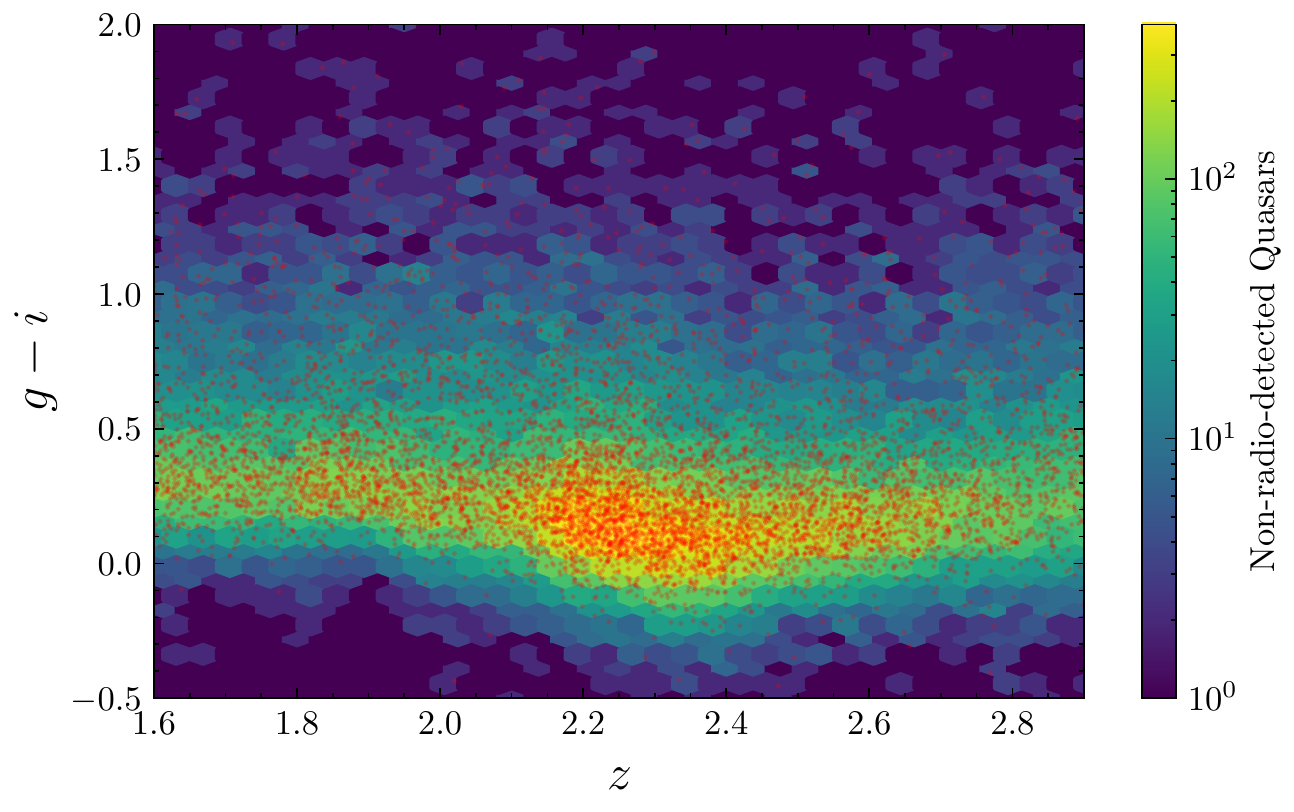}
    \caption{The colour-redshift distribution for all quasars in DR12 in the background with the radio matched sources overlayed on top in red.}
    \label{fig:colour_z}
\end{figure}

Recent work on red and blue quasars has shown similar radio-detection fraction differences between quasar populations, using LoTSS and FIRST, that are driven by optical/UV colour and support an evolutionary hypothesis for the appearance of red quasars \citep{Klindt2019FundamentalOrientation, Rosario2020FundamentalLoTSS, Fawcett2020FundamentalQuasars}. We investigate what impact the colour of our sample has on the detection fraction of BALQSOs. However, it is known that colour can have a strong dependence on redshift that should be factored in when trying to compare the colour change across a range of redshift. We use $g$ and $i$ band colours, consistent with work on red quasars, in this analysis. The overall trend in $g-i$ colour and redshift is shown in \autoref{fig:colour_z}. We limit to a redshift range $z<3$ as beyond this point the Lyman break starts to affect the colour measurement. 

\begin{figure}
    \centering
    \includegraphics[width = 1 \linewidth]{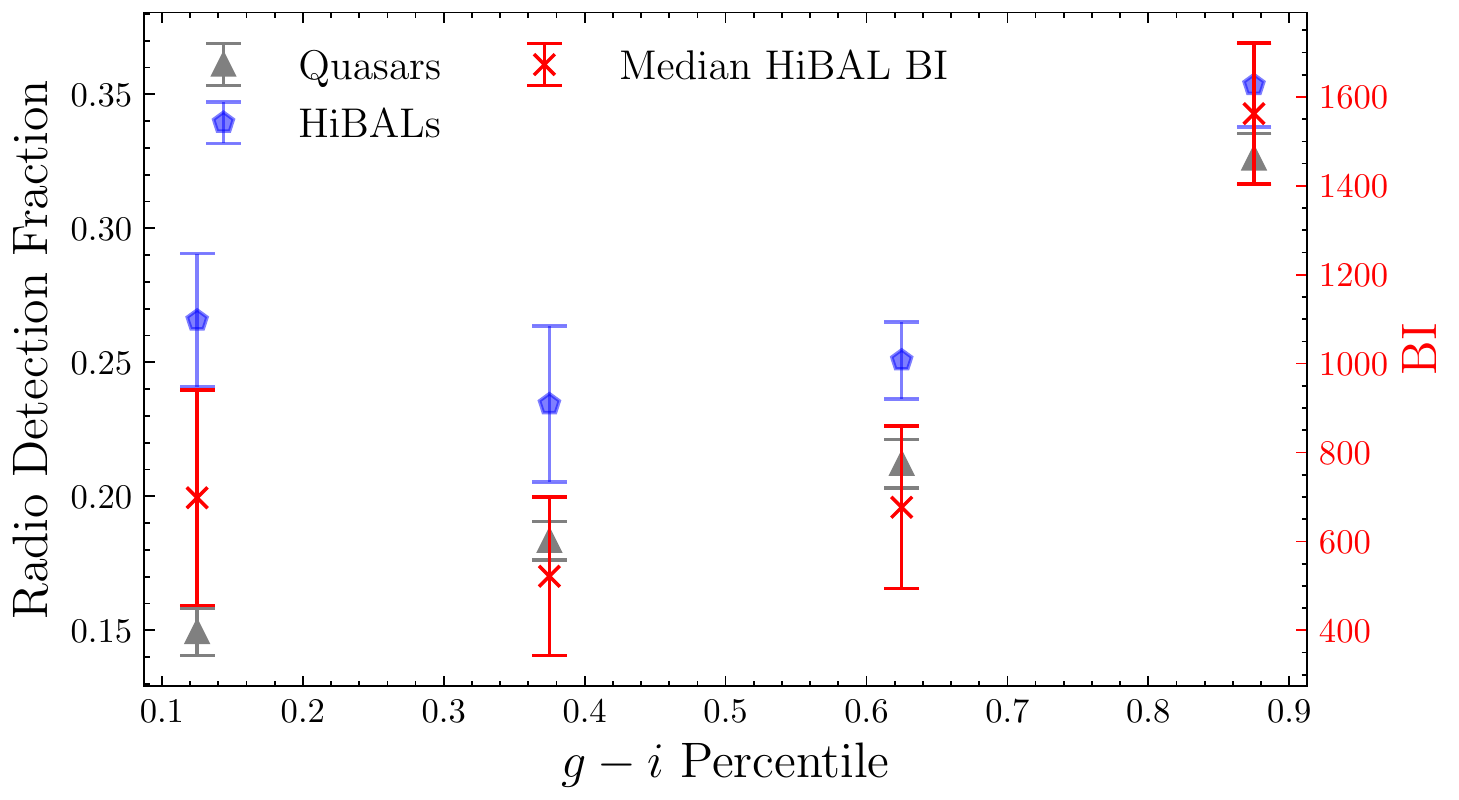}
    \caption{For each colour quartile for our quasar sample we split into general quasars and HiBALs across redshift. The colour quartiles are defined by the whole population. The median BI for HiBALs in each quartile is shown by the red dashed lines. The overall effect of this colour change from the lowest to highest quartile is close to 20\% for non-BALQSOs, close to that of the change from lowest to highest BI in \autoref{fig:frac_with_BI}. }
    \label{fig:colour_with_frac}
\end{figure}

It is clear that the redshift change should be accounted for so that we can properly determine the effect of colour on radio-detection. In \autoref{fig:colour_with_frac} we show the effect on radio-detection fraction of colour. We bin our sample first into ten equally spaced redshift bins and then into quartiles by colour in each redshift bin. This means that HiBALs and non-BALQSOs are grouped by the same bin edges. Then we average the detection fraction for each quartile across redshift. This allows for the overall trend in detection fraction with colour to be viewed while removing the redshift effects clearly visible in \autoref{fig:colour_z}. We plot the median BI for HiBALs in each bin and the scale is shown on the y-axis. The change in median BI for each percentile bin shows that the radio-detection fraction of bluer (low $g-i$) HiBALs differs from that of non-BALQSOs.

\subsection{Composite Spectra} \label{composite_results}

We use composite spectra to compare the continuum and absorption properties of a variety of sources contained within our sample. Composite spectra allow for the characterisation of large populations and also provide a way of visually identifying potential areas of further study in the different properties of the detected and non-detected quasars in our sample. 

To generate composite spectra we download spectra in batches from SDSS. We then shift these to rest wavelength and normalise by taking the median over a continuum emission range of $2575\Angstrom < \lambda < 2625\Angstrom$. This region is covered by all quasars in our sample from redshift $1.7<z<4.3$ and contains no strong emission or absorption features. This means that the flux scale is now in arbitrary units where all spectra have a value of 1 at their median point in this range. This removes effects due to the differing flux of each source and, since this point is not close to strong emission lines, it provides a good way of viewing the relative colour of different populations. Then we take a one dimensional interpolation of the spectra using the \verb|scipy interp1d| function. We can then map this interpolation to a common wavelength grid for all of the spectra. Finally, we take the median flux value at each point along the common wavelength grid to produce a final composite. We bootstrap sample the normalised spectra to obtain an uncertainty estimate on the median composite which we show as the shaded region in all composite spectra plots. One bootstrap sample consists of randomly taking 60\% of the spectra for a composite with replacement and determining the median. The uncertainty is calculated as the standard deviation of the median after 500 bootstrap samples multiplied by the square root of 60\%.

\begin{figure}
    \centering
    \includegraphics[width = 0.95\linewidth]{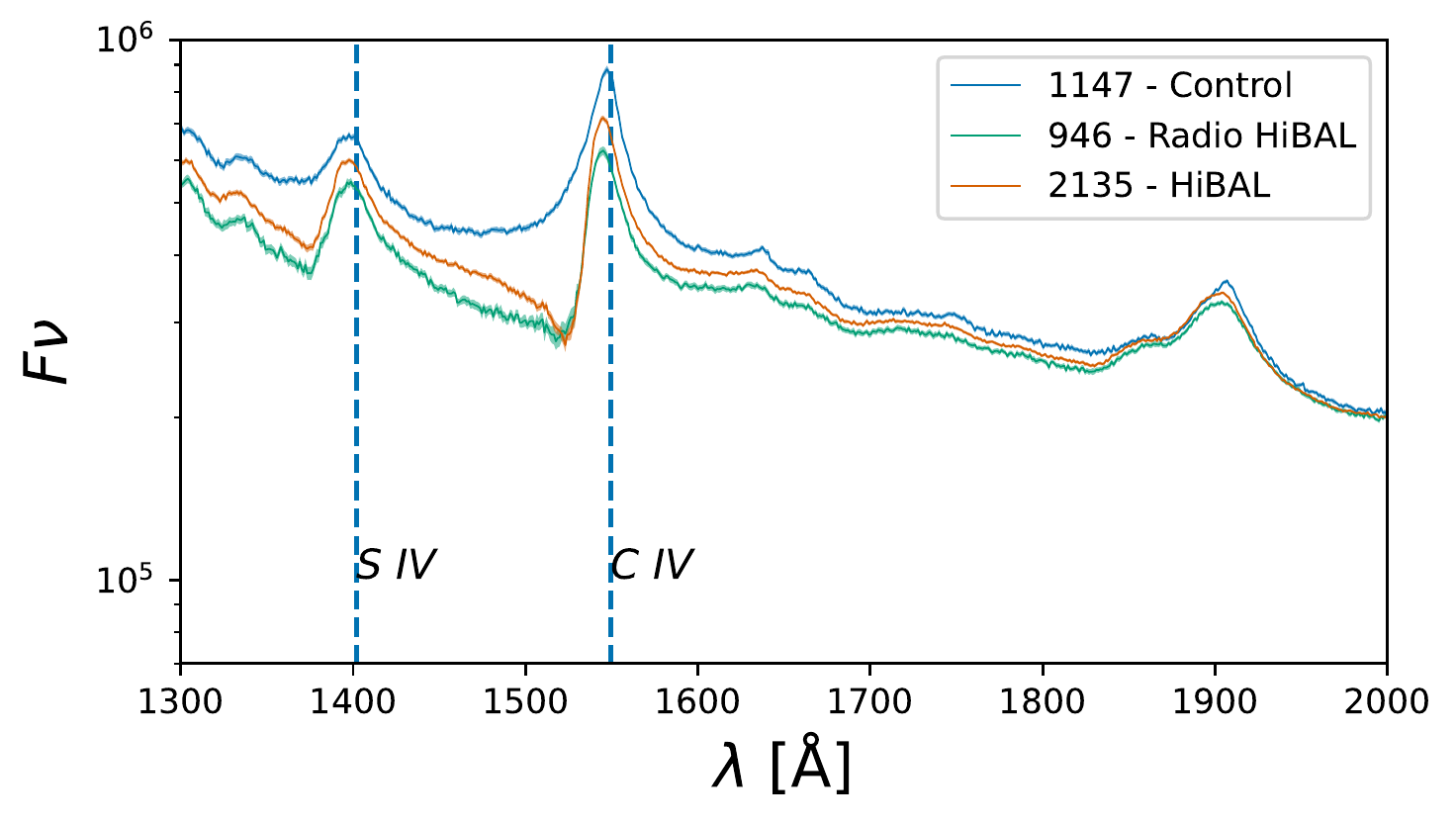}
    \caption{Overall composite for radio (green) and non-radio-detected (orange) HiBALs along with a blue quasar composite. The radio composite shows increased reddening compared the non-radio and also an absorption profile that shows more absorption at lower wavelengths but lacks the narrow feature of the non-radio composite close to C~IV.}
    \label{fig:overall_hibal}
\end{figure}

\begin{figure}
    \centering
    \includegraphics[width=0.95\linewidth]{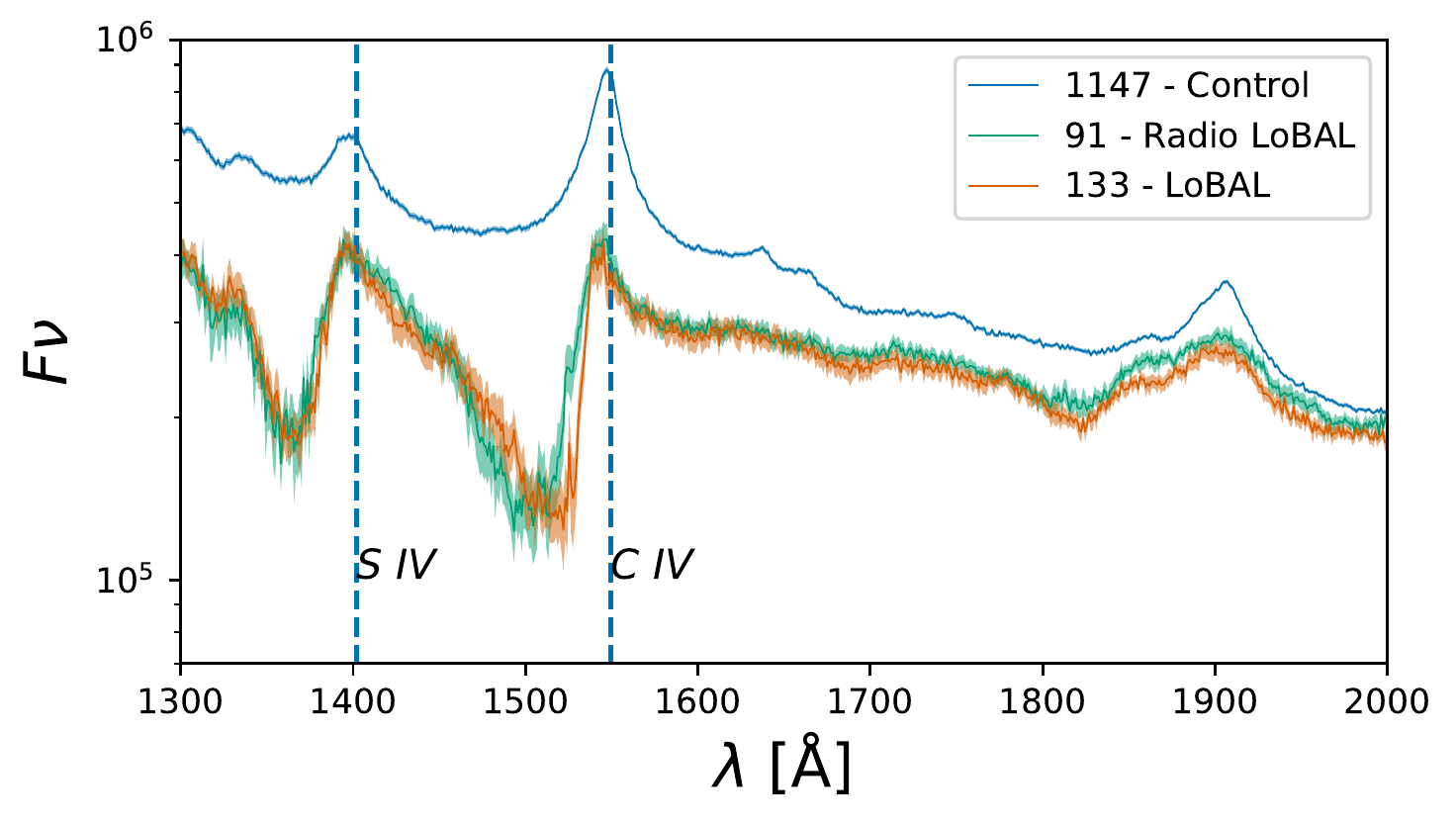}
    \caption{Overall composite for radio (green) and non-radio-detected (orange) LoBALs along with a blue quasar composite. The radio composite is less reddened compared the non-radio, a behaviour that differs to the HiBAL populations. The radio composite also shows a small shift in the base of the absorption trough towards lower wavelengths.}
    \label{fig:overall_lobal}
\end{figure}

To quantify the different amounts of extinction present in our composite spectra we create a control composite of blue quasars. This was made by defining blue quasars to be in the 25th-75th percentile range for $g-i$ colour in each redshift bin we take between $1.7 < z < 3.0$. This creates a large sample which when stacked gives a high quality composite spectra that has a minimal amount of reddening,  with a low fractional standard error at each wavelength point. It is important to note that this is not an entirely "de-reddened" spectra as typical quasars will still contain a modest amount of dust (Fawcett et al. submitted). We use it as a basis to compare the reddening of other composites.

The composite spectra we create can be used to compare continuum, emission and absorption properties when grouping the sources into meaningful samples. We construct composites based on several factors that we know connect with radio-detection fraction. This involves HiBAL and LoBAL sub-classes, BI and radio luminosity. 

Composites created using all the radio-detected and non-detected HiBALs and LoBALs are shown in \autoref{fig:overall_hibal} and \autoref{fig:overall_lobal} along with the control composite. We see that as a whole, both populations show increases reddening as expected. Radio-detected HiBALs show further reddening than the non-detected HiBALs while both LoBAL subsets are more effected by extinction but without a difference between each other. However, these overall differences are largely expected considering that we have shown how radio-detection correlates with both increased reddening and BI and these differences are reflected in the composite spectra.

\begin{figure*}
    \centering
    \includegraphics[width = \textwidth]{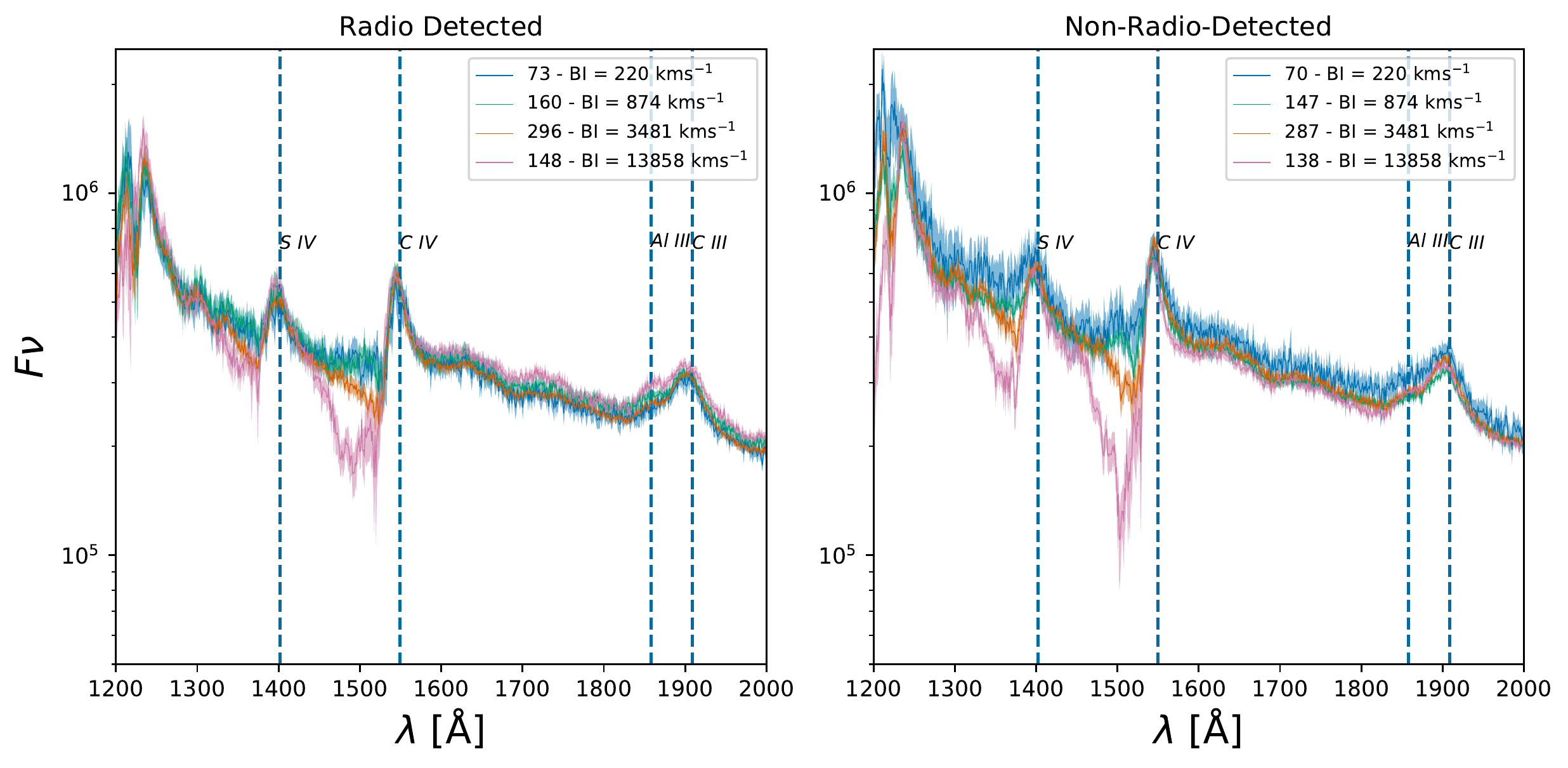}
    \caption{The reddening connection to BI, an attempt to measure wind "strength", differs for radio-detected sources compared to the general increases in reddening with BI for non-radio sources. We also demonstrate how radio BALQSOs have broader absorption features, suggesting that the radio emission could be connected with the velocity of the outflow.}
    \label{fig:BI_range}
\end{figure*}

We test the difference in composite spectra between radio and non-radio BALQSOs when matching in BI, to see if there are any changes to the absorption features and reddening with increasing wind strength. We take the HiBALs in our sample and split into 4 BI bins in log-space.  By doing this we are trying to remove effects of wind "strength" and isolate the difference in spectra that may be connected to the radio emission. In \autoref{fig:BI_range} we present the changing spectral features for radio-detected and non-detected HiBALs with BI. In general the radio-detected BALQSOs are more reddened as expected. The radio-detected BALQSOs have broader absorption features which are particular apparent in the highest BI bin.  

\begin{figure}
    \centering
    \includegraphics[width = \linewidth]{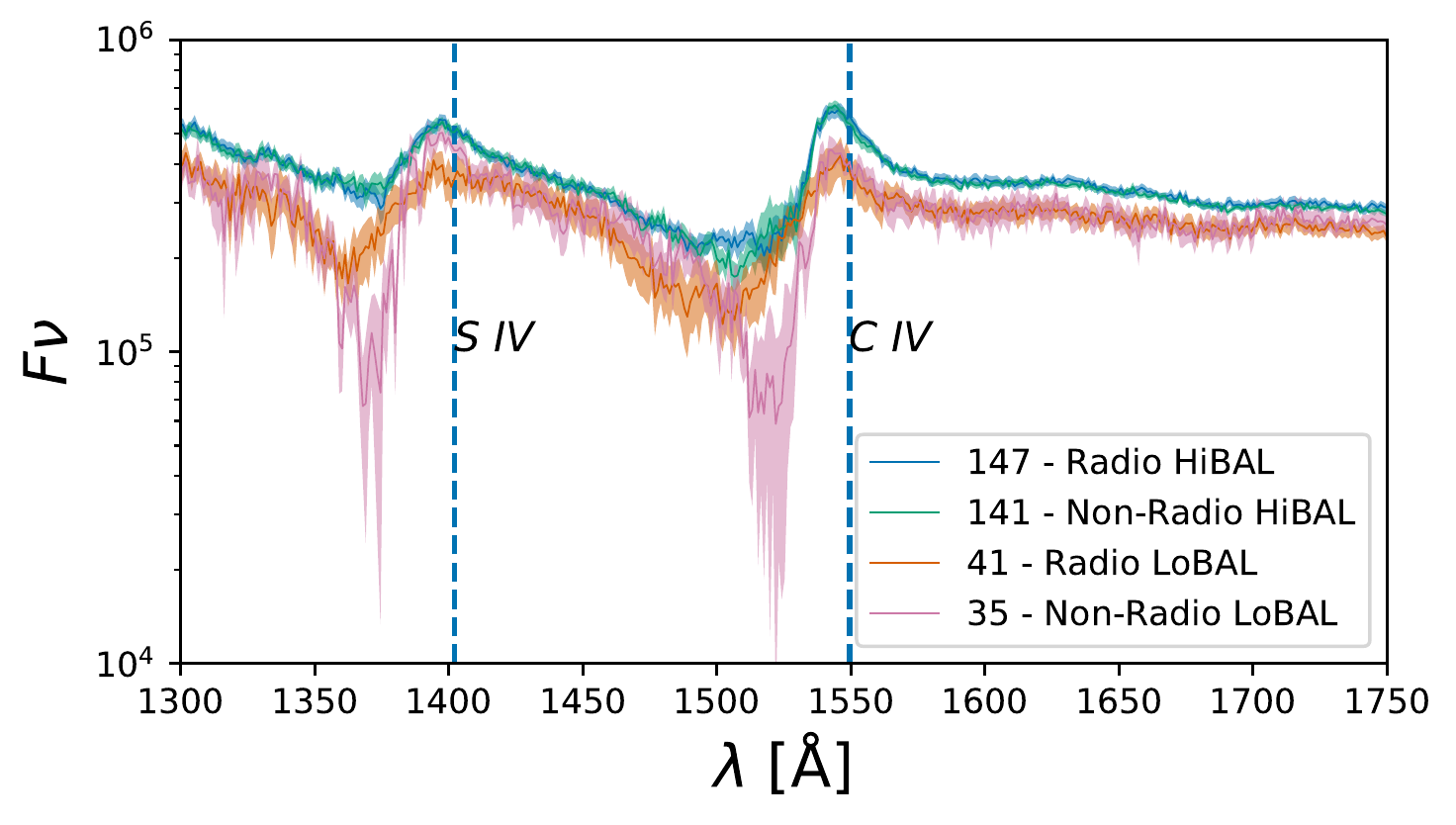}
    \caption{Composite spectra for HiBALs and LoBALs in the BI range of 3500 to 7000 kms$^{-1}$. Both LoBAL spectra show increased reddening compared to the HiBALs but they differ in their absorption properties. The radio LoBAL shows a wavelength shift in the absorption feature.}
    \label{fig:matched_hibal_lobal}
\end{figure}

Next, we compare HiBALs and LoBALs while matching in BI to remove the previously indicated effects on HiBALs of increased reddening and radio-detection at higher BI values. In \autoref{fig:matched_hibal_lobal} we match the BI range of the HiBALs and LoBALs to 3500--7000 kms$^{-1}$ and match the distributions within 5 smaller sub-bins of uniform width in that range. This gives radio and non-radio composites for HiBALs and LoBALs over the BI range. The two resulting HiBAL composites are very similar which is expected since they are at the high BI end of their distribution and in \autoref{fig:BI_range} we show that differences in reddening between the radio and non-radio HiBALs spectra become indistinguishable at high BI. The HiBALs are less reddened then the LoBALs at the same BI. This could have implications for the dust fraction in LoBALs since if the dust causing the reddening is associated with the wind or the wind strength then this alone cannot explain the reddening differences as we have matched in BI and the reddening differences between HiBALs and LoBALs remain the same. 

\begin{figure}
    \centering
    \includegraphics[width =\linewidth]{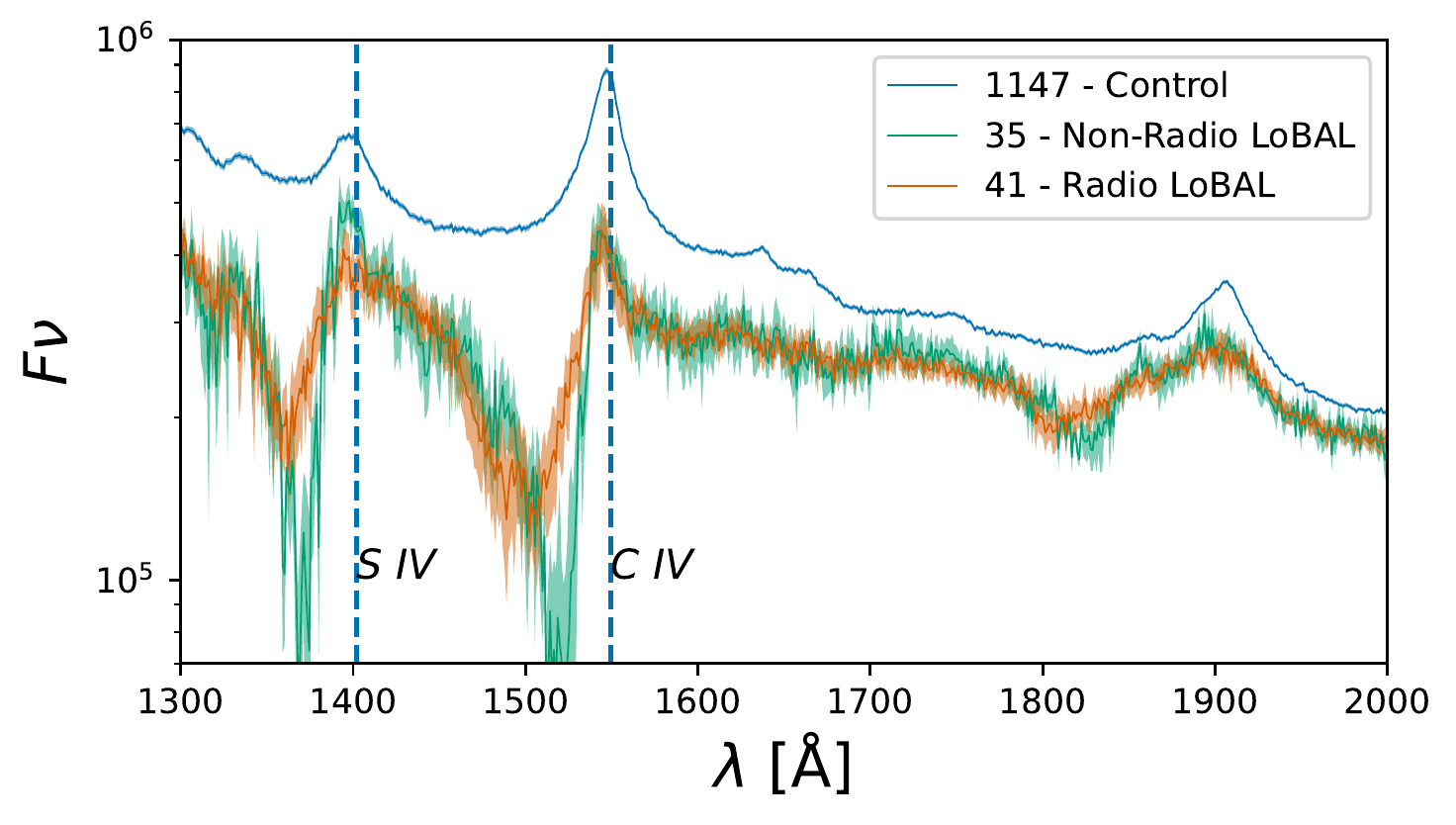}
    \caption{Both LoBAL populations show significant reddening compared to the overall control quasar composite. However, their appears to be no difference in the attenuation between the radio and non-radio-detected samples. Radio-detected LoBALs show an earlier saturated absorption trough. Saturated BAL troughs are usually explained either through partial covering of the continuum source or through scattering from an opacity not within the wind.}
    \label{fig:matched_lobal}
\end{figure}

Overall the shape of the absorption features in \autoref{fig:matched_hibal_lobal} are similar for HiBALs and LoBALs, although there is a wavelength shift in the absorption base of the radio LoBALs. This can be quantified using the relative velocity of the highest absorption (lowest flux) point of the absorption trough relative to the \ion{C}{iv} emission line. For the non-radio HiBAL the velocity shift is 5400~kms$^{-1}$ while for radio HiBALs it is 5600~kms$^{-1}$. There is an order of magnitude change when looking at LoBALs however, implying a strong connection between the wind and radio emission in these sources. For non-radio-detected LoBALs the shift is 6000~kms$^{-1}$ while for radio-detected LoBALs this value is 7100~kms$^{-1}$. We emphasise this change by showing only the LoBAL composites at matched BI, along with a control composite, in \autoref{fig:matched_lobal}. 

LoBALs show significant reddening compared to control quasars, as has been seen in other studies of this population \citep{Urrutia2009TheQuasars}. Interestingly, there is no difference in the reddening between the radio and non-radio composites. This contrasts with the HiBALs where small differences were seen as BI changes, although this LoBAL sample is already at a higher BI due to the nature of the BI distribution of LoBALs and HiBALs (see \autoref{fig:BI_dist}). In the HiBAL composites we saw that the reddening differences are not so apparent in the high BI bin. Splitting LoBALs into BI bins greatly increases uncertainty due to their rarity making it difficult to test if the same trend is present in the LoBAL population.

\subsection{\ion{C}{iv} Emission Space}

Recent work focusing on the \ion{C}{iv} emission properties for SDSS quasars has found some important connections between key parameters in this space and radio-detection. This space was initially identified in \cite{Richards2011UnificationEmission} where \ion{C}{iv} blueshift was found to larger in radio-quiet quasars as compared to radio-loud and that this could be connected to a model in which radio-loud quasars are disk dominated systems while radio-quiet quasars are wind dominated. \cite{Kratzer2015MeanEmission} also finds that the radio loudness fraction of quasars decreases as the presence of strong radiation line-driven winds increases. Most recently \cite{Richards2021ProbingQuasars} finds that the radio detection fraction is a non-linear function of \ion{C}{iv} emission properties. They argue that their sample of radio-quiet quasars may require multiple multiple radio emission mechanisms to fully explain their radio observations with the VLA and LOFAR. 

The most direct comparison to the sample in this paper can be found in \cite{Rankine2021PlacingFormation} where the LoTSS radio-detection of quasars at a similar redshift range from SDSS DR14 was found to connect to the \ion{C}{iv} emission and specifically that it correlates with \ion{C}{iv} blueshift. SDSS quasars with no \ion{C}{iv} blueshift having a radio-detection fraction of 12\% while those with a blueshift of 3000kms$^{-1}$ having a radio-detection fraction of 40\%.

An earlier study, investigating the properties of BALQSOs in this space, also outlined the method of using mean-field independent component analysis to create spectral reconstructions for each quasar \citep{Rankine2020BALSample}. They found that the fraction of BALQSOs increases with \ion{C}{iv} blueshift and lower EW. 

We wanted to test whether the radio-detection fraction behaviour we have observed so far can be explained by the \ion{C}{iv} emission space trends and whether a connection between BI and blueshift is apparent and can explain the correlation we see between BI and radio-detection fraction. To that end, we use the same method as \cite{Rankine2020BALSample} and \cite{Rankine2021PlacingFormation} to create spectral reconstructions of our composite spectra and place them in \ion{C}{iv} EW and blueshift space. 

We find that low BI BALQSOs occupy a similar blueshift to the control composite but at lower EW. Then as BI increases the blueshift also increases but that this alone is not enough to explain the radio-detection fraction increase with BI for BALQSOs. Our highest BI bin we create a composite for has a blueshift of around 1500kms$^{-1}$ for the non-radio-detected sample and 1900kms$^{-1}$ for the radio-detected sample. The radio-detection fraction for BALQSOs at this point is close to 50\% while for quasars in general with the same blueshift, \cite{Rankine2021PlacingFormation} finds the radio-detection fraction to be 25\%. BI and blueshift correlate with one another but blueshift alone cannot explain the correlation between BI and radio-detection fraction.

\section{Discussion}

Overall we find that BALQSOs show an increased LoTSS detection fraction compared to the general quasar population, consistent with previous studies, but further to this we show that these differences still persist when accounting for other known correlators with radio-detection fraction such as colour, bolometric luminosity and \ion{C}{iv} emission properties. When comparing the sub-populations of HiBALs and LoBALs we find an increased detection fraction of LoBALs as compared to HiBALs. We show that the correlation of both BI and optical reddening is not enough to explain the detection fraction differences of HiBALs and LoBALs.

We return to the SDSS data to create composite spectra for important sub-samples of quasars, HiBALs and LoBALs to investigate how the absorption profiles of these groups change depending on whether they are detected in LoTSS, on their wind strength measured by BI and on their colour. We find that radio-detected BALQSOs tend to show different \ion{C}{iv} absorption, with broader absorption features compared to non-radio-detected BALQSOs, and that these differences are most clear at the highest BI values. We also show that at matched BI the absorption profiles of radio and non-radio LoBALs show a difference in the wavelength at which absorption reaches a maximum. 

In this discussion we aim to present different ways in which to link the intrinsic radio-detection enhancement of BALQSOs, and LoBALs, with the different spectral features that we have identified with composite spectra.

\subsection{BI - Reddening Connection}

We turn to the results of Section \ref{composite_results}, specifically the positive correlations between BI and reddening and also between radio-detection in BALQSO and reddening. These changes in reddening are quite small but we discuss how they may be interpreted.

The physical explanation for increasing reddening with BI could be simply that the wind is the direct cause of the reddening and that the higher BI values are mainly due to more absorbing gas in the wind and this connects to an increased amount of dust in the wind. For the non-radio case the material increase appears to be at lower velocities as the main cause of the increase in BI is the low-velocity absorption features close to the \ion{C}{iv} emission line in the right-hand panel of \autoref{fig:BI_range}. For the radio sample, the absorption increases across a wider range of velocities in the high BI bin.

The trend in reddening and BI and the fact that BALQSO winds are known to be launched at scales $>$100~pc from the AGN \citep{Arav2018EvidenceSource, Choi2022TheSimBAL} support the conclusion that the reddening seen in BALQSOs is caused by the BAL wind itself. However, how to connect the radio emission to this wind is not so clear. At low BI the composite spectra of radio-detected BALQSOs are more reddened. Assuming that dust grains are similar in radio and non-radio-detected BALs, the increase in reddening observed has to be caused by an increase in intervening material, either in the galaxy as a whole or in the particular line of sight through to the BAL wind. In the latter case, the radio must also then have an angular dependence causing its appearance, implying small scale jets or radio emission from the wind itself while a galaxy wide increase in dust causing the reddening could allow for star formation. 

We do not investigate what reddening laws best fit our different BALQSO sub-classes but we note that work from \cite{Fawcett2022FundamentalX-shooter} indicates that the more extreme reddening in quasars may be better characterised by steeper extinction laws, corresponding to smaller dust grains, but this is very tentative. Under the assumption that the dust grains are similar in HiBALs and LoBALs, we can understand LoBALs as having higher strength winds, measured by BI, and having a denser ISM, as measured by the extinction, which could support increased radio emission through disk wind shocks.

\subsection{How do absorption features in LoBALs differ from HiBALs?}

The radio and non-radio-detected LoBAL composites show very different absorption features in \autoref{fig:matched_lobal} and also different changes relative to each other when compared to HiBALs. Firstly, we consider the average shape of the \ion{C}{iv} absorption features. The spectrum of the radio-detected LoBALs shows a flattening at the bottom of the trough in the absorption region starting at around $1475\Angstrom$ while the non-radio composite continues to have a deeper trough all the way up to the blue wing of the broad \ion{C}{iv} emission line. Determining whether the depth of a BAL trough is the result of increased column density, 'non-black' saturation caused by partial covering of the emission source \citep{Arav1999What1603+300}, scattering from another source that is not co-spatial to the wind \citep{Lamy2000Spectropolarimetry0059-2735} or from the combination of several wind components at different velocities is an outstanding problem in the analysis of these winds. Whatever the cause of  potential saturation, it more strongly effects the radio-detected BALQSOs.  

\begin{figure}
    \centering
    \includegraphics[width = \linewidth]{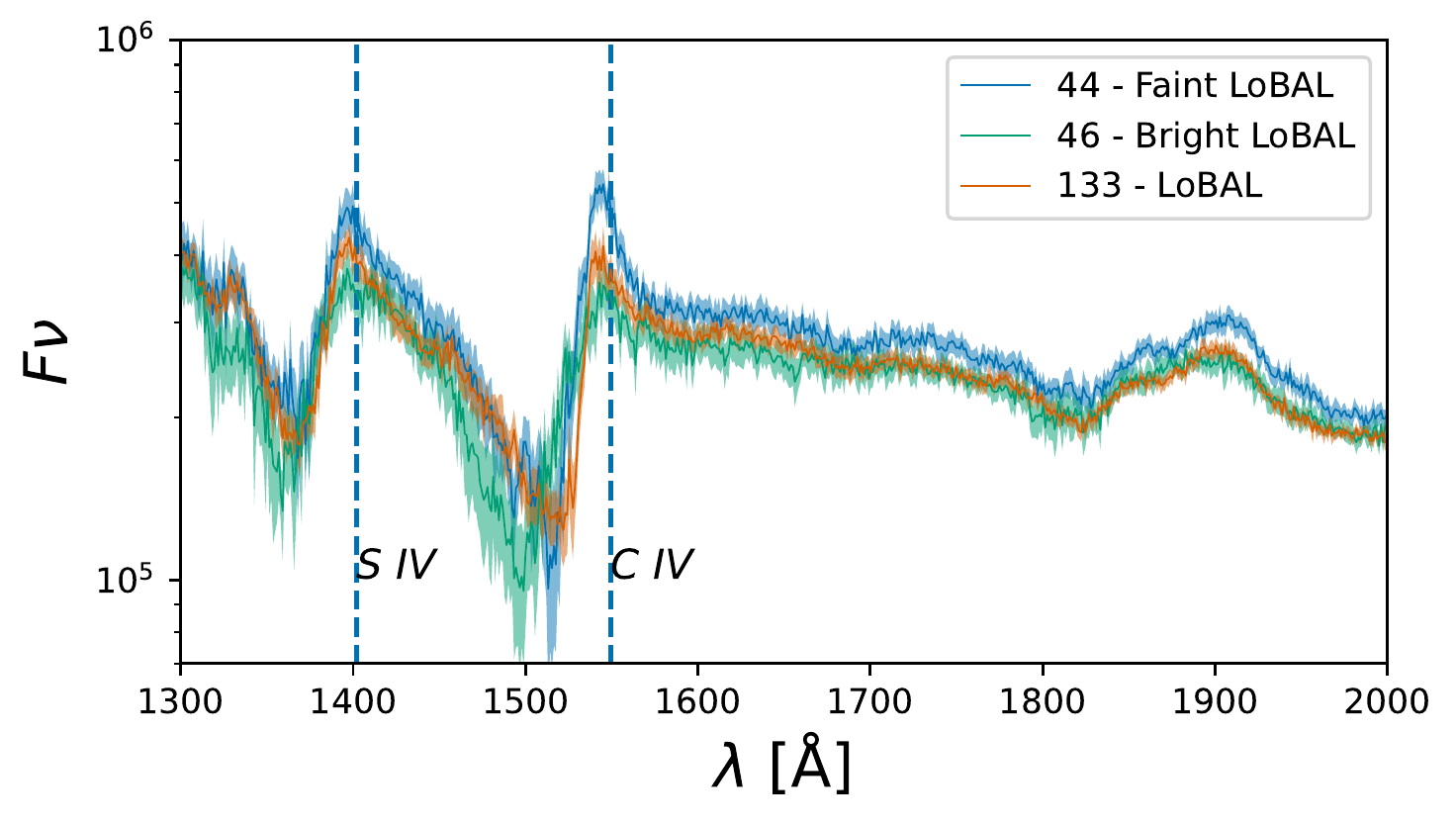}
    \caption{By splitting into a high and low radio luminosity sample, by the median LoBAL radio luminosity, we examine the dependency of absorption properties on the magnitude of the radio emission. We find connections between the velocity of the outflow and the radio emission. The blue-shift of the radio composite implies higher velocity outflows along with an increased amount of intervening material responsible for the saturation and increased reddening in the radio composite.}
    \label{fig:lobal_brightness}
\end{figure}

Further evidence for this effect can be found when splitting the radio-detected sample into a high (bright) and low (faint) luminosity bin divided by the median BALQSOs radio luminosity ($3 \times 10^{25}$~WHz$^{-1}$ for HiBALs and $3.7 \times 10^{25}$~WHz$^{-1}$ for LoBALs). Composite spectra for these samples, along with the non-radio-detected LoBAL composite, is shown in \autoref{fig:lobal_brightness}. The faint population closely resembles the non-radio-detected population in terms of their absorption profiles and is  interestingly less affected by reddening. However, in the higher radio luminosity bin, several changes occur. Firstly, the extinction increases so that the reddening is similar to the non-detected sample. The lower wavelength of the base of the absorption trough increases compared to the other samples, implying a higher velocity for the absorbing gas in this more radio luminous sample or a greater launch radius from the central black hole. Finally, the absorption profile also appears to be broader for radio-bright LoBALs. These results do suggest velocity of the wind being a factor in radio emission, along with an increased amount of material responsible for the reddening, which could come from the wind or from other in-situ dust.

Increased reddening for the radio-detected sources combined with the higher incidence of broader absorption in the \ion{C}{iv} absorption region is potentially suggestive of increased amount of material along the path of the high-velocity outflow. This could make radio emission from shocks, as the outflow interacts with this material, a potential mechanism for the radio-detection fraction increase in LoBALs compared to HiBALs. We explore this possibility, along with alternative explanations for radio emission, in the rest of this discussion.

\subsection{Radio Interpretation}

The origin of the radio emission in radio-quiet quasars is a topic of much debate and this extends to BALQSOs, an overwhelmingly radio-quiet population. Three main mechanisms have been proposed as the dominant contributor: star formation \citep{Padovani2011TheNuclei}, frustrated jets \citep{White2015Radio-quietMJy} and the disk wind \citep{Blundell1999TheSamples, Faucher-Giguere2012TheNuclei, Fukumura2013StratifiedJets}. Conversely, the consensus for the radio-loud population is that radio jets are the dominant factor and this has allowed for the orientation of radio-loud sources to be probed more easily as resolving the jet components can allow inclination to be determined under certain assumptions of symmetry. Connections have been made between the wind and jet for radio-loud sources; for example \cite{Mehdipour2019RelationAGN} find a negative correlation between radio-loudness and the column density of the ionised wind. They attribute this coupling to magnetic mechanisms powering both the winds and the radio-loud jet, providing a connection between the two and originating at the black hole scale. This connection is complicated by time scale being an important factor when considering radio-loud sources. The radio-loud emission attributed to large jets may take $\sim10^7$ years to evolve \citep{Hardcastle2018AGalaxies}, while the magnetic field and wind column density evolve at smaller size scales and change on much faster timescales. Radio-loudness is not a measure of the current state of the black hole and will continue to increase if a jet can continue to be powered over a long period of time while the optical continuum emission, part of the loudness ratio, can change much more quickly as the accretion close the black hole is on a much smaller scale.

The connection between BI and radio-detection fraction presented in the previous section (see \autoref{fig:frac_with_BI}) has a key role to play when attempting to constrain the source of the emission and the structure of BALQSOs. For example, if one takes the evolutionary theory for BALQSO occurrence, that BALQSOs are an early phase in the lifetime of a quasar in which much of the star formation quenching is triggered, and that jets are the radio source, then the connection with high BI and radio-detection fraction appears hard to explain as because the UV line-driven wind and the majority of radio originate on totally different size scales they also differ in time by a large amount. The wind and jet connection from \cite{Mehdipour2019RelationAGN} cannot be due to synchronous effects, although the overall history of the AGN could be the connection. Studies of disappearing and appearing BALQSOs \citep{DeCicco2017CSurveys, Mishra2021AppearanceQuasars} have placed the variational time scale for BALQSO winds, on the order of hundreds of years, to be much shorter than that of radio jets, on the order of Myrs. Therefore the BI to detection fraction correlation lacks the synchronicity one might expect in the evolutionary theory. This could be resolved by invoking a number of BALQSO phases, thereby extending the overall time in which a BALQSO wind could be observed and connecting these phases with the intermittent power changes that are known to occur in the production of radio jets around black holes \citep{Nyland2021VariableSurvey}.

Overall, we have established that both BI and colour correlate with radio-detection fraction but since these two parameters are likely connected it is hard to tell what is really driving the detection fraction increase. The absorption profiles of radio-detected BALQSOs show lower-wavelength absorption and also reduced narrow line absorption close to the \ion{C}{iv} line when compared to the non-radio-detected BALQSOs, particularly at higher BI values. When we match in BI for both the HiBAL and LoBAL populations, we find differences in the radio-detected and non-radio-detected LoBAL populations but not for HiBALs. 

We know from \autoref{fig:radio_vs_bol} that HiBAL and LoBAL populations have similar distributions in radio and bolometric luminosity so the emission mechanisms are likely to be similar in both populations. The fact that at matched BI, HiBALs and LoBALs show different absorption feature changes depending on whether they are radio-detected or not, suggests that that an orientation model cannot explain the transition from HiBAL to LoBAL. 

Ultimately, the radio-detection fraction differences between HiBALs, LoBALs and non-BALQSOs must be driven by one of two general mechanisms: either star formation or processes associated with the AGN. We now focus on each of these in more detail.

\subsection{Radio Interpretation - Star Formation}

We show clearly that the wind properties and radio emission are connected via our composite spectra, immediately making a star formation explanation for the radio detection fraction enhancement unlikely. Also the correlation between BI and radio-detection fraction is hard to interpret in a star formation scenario as the increased wind-strength would need to couple to triggering bursts of star formation. 

A case could be made for an indirect link via the fact that star formation and AGN accretion relies on the same fuel source. Maybe this could connect the BALQSO wind appearance to increased star formation. However, this would rely on a similar timescale for the fuelling of both the star formation and the BALQSO wind. Studies have shown that these timescales are not at all similar for star formation and other forms of AGN activity \citep{Hickox2014BlackNucleus}, implying that this would also be the case for BALQSO winds. 

Star formation as the primary radio emission mechanism for BALQSOs is not compatible with the angular interpretation for the presence of BAL emission troughs. Simply put, radio emission from galactic star formation should have no dependence on viewing angle. In general, after finding the strong radio-detection differences in BALQSOs, to maintain an orientation based approach one must also find a radio emission mechanism that is also orientation dependent as the cause, or one must involve relativistic beaming of the proposed mechanism at an angle that relates to the angle of the BALQSO wind.  

Another way to interpret the increased detection fraction of BALQSOs, relative to the general population, is to actually consider the possibility of there being a deficit in flux density from typical quasars due to increased free-free absorption at these LOFAR frequencies. This would then be difficult to explain with either the orientation or evolutionary scenarios. 

In the evolutionary explanation of BALQSO occurrence, LoBALs are typically placed at an early stage in the evolution of a quasar. Therefore it is expected that LoBALs may have higher star formation rates as they are more reddened and potential quenching effects of the outflows and quasar more generally have had less time to take effect. The powerful winds in BALQSOs could then remove the dust associated with star formation and leave a blue quasar behind.  

Previous studies have tried to find differences in the IR and line ratio properties of LoBALs and HiBALs that could suggest that they have different star formation rates to the general population. Recent examples such as \cite{Wethers2020Star2.5} and \cite{Chen2021EvidenceQuasars} both claim to find evidence that LoBALs show signs of being in a special phase of high star formation which is not seen in non-BALQSOs or HiBALs. However, these were certainly not conclusive findings and a larger study of HiBALs and LoBALs using more accurate star formation tracers with an instrument such as ALMA will be needed to test these results. 

If star formation is the cause of enhanced radio-detection in HiBALs and then even further in LoBALs, then the differences between their composite spectra ought to remain consistent with this explanation. However, we find that radio LoBALs show different absorption features at matched BI to radio LoBALs. This would suggest that the wind properties cannot be related to the star formation that is causing the radio-detection fraction increase for BALQSOs compared to non-BALQSOs. The only resolution would be that both the BALQSO wind appearance and enhanced star formation have an evolutionary explanation that is separate for both but typically occurs on the same cosmological timescale giving rise to the radio-detection fraction increase for BALQSOs. 

Generally, the energetic output of winds is thought to likely reduce star-formation through the heating of molecular gas clouds which are much more likely to collapse and form stars when they are cooler. However, we do consider scenarios in which the presence of an outflow may in fact trigger star formation episodes, likely through the compression of the molecular star-forming gas and increasing subsequent star formation. 

If star formation is the largest source of radio emission in BALQSOs then the trend in BI and radio-detection fraction would suggest that stronger winds can trigger more star formation, although it is important to note that no correlation was found between BI and the radio luminosity itself. The timescale of both the wind lifetime and the time needed to trigger a significant amount of star formation are important when considering whether the direct connection is plausible. Star forming clouds collapse on timescales on the order of Myrs to tens of Myrs. This is potentially much larger than the lifetimes of a BAL wind, especially when considering there is likely a delay between the hypothetical compression of the star forming gas clouds and the time at which the star formation levels are high enough to drive a significant increase in the radio-detection fraction.

Even if there is an increased level of star formation, any radio emission from an AGN process could be added on top of the star formation level, potentially weakening observed trends with radio luminosity.

\subsection{Radio Interpretation - AGN}

AGN-related radio-quiet emission mechanisms include frustrated sub-galaxy scale jets, weak jets, wind emission and coronal emission (\citealt{Panessa2019TheNuclei} and references therein). We can largely rule out coronal emission since the radio-detection fraction correlates with larger blueshifts which are not connected. In this subsection we take each mechanism in turn and, assuming that the mechanism is the largest contributor to radio emission in BALQSOs, consider that process' plausibility given the results of this work.

\subsubsection{Frustrated Jets}

Frustrated jets are used as an explanation for radio-quiet emission where the jet structure is similar or identical to that of radio-loud and FR1 type sources, but they have a different efficiency in accelerating electrons \citep{Falcke1995TheQuasars.}. The clear signature for this process being the key contributor to radio emission in BALQSOs would be an excess of compact and flat spectrum sources. Using the LoTSS DR2 definition of whether a source is resolved from \cite{Shimwell2022TheRelease}, we do in fact find that BALQSOs are slightly more likely to be unresolved at 6 arcsecond resolution than non-BALQSOs in LoTSS DR2. BALQSOs are 98\% unresolved while the general population are 96\% unresolved. They also have on average flatter spectra when using LoTSS DR2 and FIRST, which was also found in DR1 \citep{Morabito2019}, with a median spectral index of $0.094 \pm 0.25$  compared to $-0.65 \pm 0.091$ for non-BALQSOs. It is rare to find BALQSOs with spectral indexes $<-0.5$ in our sample. It should be noted that the requirement of a FIRST detection to determine the spectral index biases the sample towards radio-loud sources. This could be seen as some small support for the jet hypothesis.

Jets are one possible emission mechanism that could allow for an angular interpretation of the BALQSO phenomena. The classic theory of radio jets suggest that they are launched perpendicular to the accretion disk \citep{Begelman1984TheorySources, Blandford1995TheQuasars.}. Therefore, one could interpret the increase in radio-detection with BI as a change in viewing angle, with LoBALs being at the closest angle to the jet (axis of rotation of the accretion disk) , allowing for a clearer view of the compact jet or the effects of relativistic beaming as it becomes more likely that our viewing angle intersects with the jet. However, upon closer inspection we suggest that in fact, with the standard model of winds being viewed closer to the accretion disk than the orthogonal jet axis \citep{Matthews2016TestingWinds}, this is the exact opposite of what we would expect. As our viewing angle gets closer to the accretion disk, and BI increases, we are in fact moving away from the jet axis and so would expect a decrease in radio-detection fraction. This seems to be strong evidence to rule out the combined theory of an orientation approach with radio jet emission for BALQSOs unless the orientation is that of a polar wind, as proposed by \citep{Ghosh2007}. Doppler beaming is very unlikely to help here either for similar reasons and because we are observing at low frequencies where the radio emission is lobe dominated. 

A key issue facing a possible jet hypothesis is the fact that BALQSOs are a radio-quiet population. Therefore if jets are the explanation for the enhanced radio-detection fraction in BALQSOs, we also need an explanation for why there is a deficiency in high-power radio jets as compared to the general quasar population. Either the jets are different in nature to radio-loud quasars or somehow the the jet destroys the BALQSO wind as it expands. 

Some research suggests that BALQSOs are more likely to be GigaHertz Peaked Spectrum (GPS) sources \citep{Bruni2015RestartingQuasars, Bruni2016FastSources}. GPS sources are known to be jet dominated and likely have younger jets. They display a turnover in the radio SED due to synchotron self absorption. If it is the case that BALQSOs are GPS-like sources then we would expect BALQSOs to display flatter spectral indices compared to non-BALQSO quasars when measuring at the 144 MHz of LOFAR and the 1.4 GHz of FIRST which is indeed what we find. This means the spectral indices that we measure in the BALQSO sample may not need an orientation interpretation and could lie in a phase before a radio jet is relaunched or recollimated.

\subsubsection{Disk Winds}

Disk winds have two main theorised radio emission mechanisms. The first is in the form of disk wind shocks and the second is from brehmstrahlung free-free processes. However, the latter was found to be unable to explain the X-ray and radio luminosities for a sample of radio-quiet quasars in the Palomar Green sample \citep{Steenbrugge2010RadioQuasars}. Also, the flat spectral shape of brehmstrahlung emission at radio wavelengths implies that its contribution to the LOFAR bands is minimal.  

Evidence that outflows trigger shock synchotron emission was found by \cite{Zakamska2014QuasarQuasars} where they show that [\ion{O}{iii}] velocity had a positive correlation with radio luminosity for a sample of quasars. Calculations from \cite{Nims2015ObservationalNuclei}, using a model of a homogeneous and spherically symmetric ISM, suggest that the radio luminosity of a wind with an energy of a few percent of the bolometric luminosity of a powerful AGN would be at least as great as star formation. 

The distance at which winds are typically launched in BALQSOs has large implications for their observational signatures and their energetic output into a galaxy. The fact that we cannot extract the velocity of the wind and distance from the AGN through the shift of the absorption trough allows for several interpretations of the BALQSO composites, in particular the differences between LoBAL groups in \autoref{fig:matched_lobal} and \autoref{fig:lobal_brightness}. In \autoref{fig:matched_lobal} we emphasise the difference in the absorption features of radio and non-radio-detected LoBALs despite matching in BI. Although there is no clear reddening difference the absorption trough shows a blue shift at all three marked spectral lines, or maybe just a saturation at C~IV, in the radio composite. This could potentially imply that radio emission is connected to the velocity of the outflow in BALQSOs. Or alternatively the radio LoBAL winds are launched at greater distances from the central black hole. In \autoref{fig:lobal_brightness} we also see a strong difference in the reddening where the brighter LoBALs have increased attenuation. Together with the absorption trough shift, it could imply that radio emission is connected to higher velocity winds or simply to a more dense ISM where the shocks are triggered. 

The saturation of features is also seen at the high BI end for HiBALs in \autoref{fig:BI_range}. Considering that LoBALs are at the high BI end of the overall BALQSO distribution it could be possible that the saturation features have the same underlying physical reason for HiBALs and LoBALs but we do not find the same shift in the absorption features. This feature is only seen when looking at the radio and non-radio LoBALs and noticing that it is more apparent at the high radio-luminosity end.

The fact that we observe changes in the absorption properties when looking at the radio detected and non-detected samples strongly implies a connection between the BALQSO outflow and the radio emission.

\section{Conclusions}

In this study we have created the largest sample of radio-detected BALQSOs by matching the latest LoTSS DR2 release to the SDSS DR12 quasar catalogue. By creating a large sample of BALQSO with high SDSS signal to noise ratio to remove certain biases in the identification of BALQSOs, we have been able to approach this population on a larger statistical scale, even for the rarer LoBALs, than any previous radio study allowing us to achieve the following key results

\begin{enumerate}
    
    \item The overall radio-detection of HiBALs and LoBALs across redshift is 30.2\% and 44.9\%. For general quasars the fraction is lower at 20.2\%. 
    
    \item The radio-detection fraction of BALQSOs has a weak correlation ($r = 0.9609$, $p=5\times10^{-4}$) with balnicity index even when excluding LoBALs which account for many of the highest BI sources. This connects the absorption line properties of BALQSOs to their radio emission. 
    
    \item Radio-detection fraction for BALQSOs also correlates with $g-i$ colour but the behaviour of the bluest BALQSOs, entirely HiBALs, differs from the general quasar population. This subgroup will need further investigation. Also BALQSOs still show an intrinsic increase in detection fraction compared to non-BALQSOs across colour space.
    
    \item Radio-detection fraction also correlates with \ion{C}{iv} blueshift in quasars. However, the BALQSO radio-detection fraction increases with BI at a rate beyond the expected change from \ion{C}{iv} blueshift increase. 
    
    \item Composite spectra of HiBALs shows that the radio-detected HiBALs are generally more reddened across BI but that this difference is reduced as BI increases. These reddening changes are small. Radio-detected HiBALs show more saturation in the absorption troughs at high BI. 
    
    \item HiBALs and LoBALs show different reddening and absorption features even when matched in BI. They also show differences in the relation between their radio and non-radio-detected spectra with the HiBAL composites being nearly indistinguishable.
    
    \item The radio-detection fraction increase for LoBALs cannot be explained through relativistic beaming of jets that are also present in HiBALs, without a polar model for all BALQSOs.

    \item Radio-detected LoBALs show a shift of around 20\%, compared to non-radio detected LoBALs, in the base of the absorption trough at C~IV. This implies either a connection between the velocity of the BAL wind and the radio emission or in the radius of launch from the AGN. Each possibility leads to several different interpretations depending on which radio emission process is assumed to be primary.

\end{enumerate}

Overall, the results from this paper point towards the radio emission of BALQSOs being connected directly to the wind, and hence likely to the radio-emission of wind shocks in the ISM being the cause for the increased radio-detection fraction. The main evidence for this comes from the absorption property changes, shown in the composite spectra, in sources that are radio-detected and from timescale issues for other explanations of radio emission with these shorter term wind changes in mind. However, we cannot entirely rule out any of the radio emission mechanisms or favour a particular orientation or evolution model.

By utilising the sub-arcsecond resolution imaging capabilities of the LOFAR international stations \citep{Morabito2021Sub-arcsecondPipeline} we intend to to characterise the morphology of many of these compact BALQSOs by using international LOFAR stations to reach sub-arcsecond resolution. This means we will be able to see if the emission has structure (galaxy scale jets), is distributed across the galaxy (star formation) or if it is still largely compact we can look at disk wind emission or frustrated small scale jets. Combining this with e-MERLIN observations to extract spectral indices at similar resolution for multiple component sources, we aim to be able to confirm which mechanisms are occurring in these sources.

\section*{Acknowledgements}

LOFAR data products were provided by the LOFAR Surveys Key Science project (LSKSP; \url{https://lofar-surveys.org/}) and were derived from observations with the International LOFAR Telescope (ILT). LOFAR (van Haarlem et al. 2013) is the Low Frequency Array designed and constructed by ASTRON. It has observing, data processing, and data storage facilities in several countries, which are owned by various parties (each with their own funding sources), and which are collectively operated by the ILT foundation under a joint scientific policy. The efforts of the LSKSP have benefited from funding from the European Research Council, NOVA, NWO, CNRS-INSU, the SURF Co-operative, the UK Science and Technology Funding Council and the Jülich Supercomputing Centre.

The Jülich LOFAR Long Term Archive and the German LOFAR network are both coordinated and operated by the Jülich Supercomputing Centre (JSC), and computing resources on the supercomputer JUWELS at JSC were provided by the Gauss Centre for Supercomputing e.V. (grant CHTB00) through the John von Neumann Institute for Computing (NIC).

This research made use of Astropy,\footnote{http://www.astropy.org} a community-developed core Python package for Astronomy \citep{Collaboration2013Astropy:Astronomy, Collaboration2018ThePackage}.

JHM acknowledges a Herchel Smith Fellowship at Cambridge. AD acknowledges support by the BMBF Verbundforschung under the grant
05A20STA.

This work was supported by grant MR/T042842/1.

This work was supported by the consolidated grant ST/T000244/1.

JP acknowledges support for their PhD studentship from grants ST/T506047/1 and ST/V506643/1.

ALR acknowledges support from UKRI grant code MR/T020989/1.

Funding for the Sloan Digital Sky 
Survey IV has been provided by the 
Alfred P. Sloan Foundation, the U.S. 
Department of Energy Office of 
Science, and the Participating 
Institutions. 

SDSS-IV acknowledges support and 
resources from the Center for High 
Performance Computing  at the 
University of Utah. The SDSS 
website is \url{www.sdss.org}.

SDSS-IV is managed by the 
Astrophysical Research Consortium 
for the Participating Institutions 
of the SDSS Collaboration including 
the Brazilian Participation Group, 
the Carnegie Institution for Science, 
Carnegie Mellon University, Center for 
Astrophysics | Harvard \& 
Smithsonian, the Chilean Participation 
Group, the French Participation Group, 
Instituto de Astrof\'isica de 
Canarias, The Johns Hopkins 
University, Kavli Institute for the 
Physics and Mathematics of the 
Universe (IPMU) / University of 
Tokyo, the Korean Participation Group, 
Lawrence Berkeley National Laboratory, 
Leibniz Institut f\"ur Astrophysik 
Potsdam (AIP),  Max-Planck-Institut 
f\"ur Astronomie (MPIA Heidelberg), 
Max-Planck-Institut f\"ur 
Astrophysik (MPA Garching), 
Max-Planck-Institut f\"ur 
Extraterrestrische Physik (MPE), 
National Astronomical Observatories of 
China, New Mexico State University, 
New York University, University of 
Notre Dame, Observat\'ario 
Nacional / MCTI, The Ohio State 
University, Pennsylvania State 
University, Shanghai 
Astronomical Observatory, United 
Kingdom Participation Group, 
Universidad Nacional Aut\'onoma 
de M\'exico, University of Arizona, 
University of Colorado Boulder, 
University of Oxford, University of 
Portsmouth, University of Utah, 
University of Virginia, University 
of Washington, University of 
Wisconsin, Vanderbilt University, 
and Yale University.

\section*{Data Availability}

The LoTSS DR2 radio data and catalogue used in this work is described in \cite{Shimwell2022VizieR2022} and can be found at the following page - \url{https://lofar-surveys.org/dr2_release.html}

The SDSS DR12 quasar catalogue used in this work is described in \cite{Paris2017VizieR2017} is located at the following URL  - \url{https://data.sdss.org/sas/dr12/boss/qso/DR12Q/DR12Q.fits}

The SDSS spectra used to create the composites can be downloaded through various means described on this page - \url{https://www.sdss.org/dr17/spectro/}.

A catalogue of SDSS quasars in the LoTSS DR2 footprint with added derived data including radio loudness and estimated rest $B$ band emission is available at CDS via anonymous ftp to cdsarc.u-strasbg.fr
(130.79.128.5) or via \url{https://cdsarc.unistra.fr/viz-bin/cat/J/MNRAS}.

Composite spectra created for the this work can be made available upon request - Primary contact is \url{james.w.petley@durham.ac.uk}.


\bibliographystyle{mnras.bst}
\bibliography{references.bib}

\bsp	
\label{lastpage}
\end{document}